\documentclass[final,2p,times,twocolumn]{elsarticle}

\usepackage[margin=1.5cm]{geometry}

\usepackage{amssymb}

\usepackage[utf8]{inputenc}
\usepackage{float}
\usepackage{amsmath}
\usepackage{siunitx}
\usepackage[version=4]{mhchem}
\usepackage{multirow}
\usepackage{subcaption}
\usepackage{placeins}
\usepackage{makecell}
\usepackage[numbers]{natbib}
\usepackage{hyperref}
\hypersetup{colorlinks,citecolor=black,filecolor=black,linkcolor=black,urlcolor=black}
\usepackage{cuted}
\usepackage{gensymb}
\usepackage{verbatim}
\DeclareSIUnit{\molar}{M}

\journal{JCIS Open}

\begin{document}

\begin{frontmatter}



\title{Colloidal structure, energy extensivity and Monte Carlo sampling properties of improved short-range interaction models for surfactant-coated magnetic nanoparticles}


\author[inst1]{Aimê Gomes da Mata Kanzaki}
\author[inst1]{Tiago de Sousa Araújo Cassiano}
\author[inst2]{João Valeriano}
\ead{joao-pedro.valeriano-miranda@univ-amu.fr}
\author[inst1]{Fabio Luis de Oliveira Paula}
\author[inst1]{Leonardo Luiz e Castro}

\affiliation[inst1]{organization={Universidade de Brasília, Instituto de Física},
            addressline={Campus Universitário Darcy Ribeiro}, 
            city={Brasilia},
            postcode={70.910-900}, 
            state={DF},
            country={BR}}

\affiliation[inst2]{organization={Aix-Marseille Université, CINAM, Turing Centre for Living Systems},
            city={Marseille},
            country={FR}}

\begin{abstract}
The standard DLVO theory offers a limited description of ionic-surfacted magnetic colloids in near aggregation regimes. Correcting the electrical double layer term for ionic surfactants is not enough to successfully simulate the systems. The correction of the van der Waals energy divergence at short interparticle distances is fundamental for proper Monte Carlo sampling of nanoparticles’ configurations. We compare different short-range interaction models and show that a more detailed model leads to Monte Carlo simulations that better match theoretical expectations. Studying the energy scaling with the number of particles, we observe a slight deviation from energy extensivity across all models, small but still detectable via Akaike’s information criterion. The more detailed model predicts a strong effect of particle-size dispersity on the transition between overall attraction and repulsion. More precise modeling can significantly affect numerical predictions, in particular, the effect of particle-size dispersity on the spatial structure of colloids with high volume fraction. This emphasizes the importance of nailing down better models for describing complex colloidal dispersions.
\end{abstract}





\end{frontmatter}



\section{Introduction}
\label{sec:introduction}

Colloids, alongside their ample occurrence in nature and everyday products, are routinely synthesized in new varieties aiming at multiple applications.
In particular, ferrofluids (magnetic colloids), made of magnetic iron oxide nanoparticles, have garnered significant interest due to their role in novel technological, biomedical, and environmental applications~\citep{Bailey-1983, Raj-1990, Roger-1999, Scherer-2005}.
Examples include their use as contrast agents in magnetic resonance imaging~\citep{Seo-2006, Harisinghani-2003}, as multifunctional agents in theranostics~\citep{Song-2020}, as a vehicle for the administration of targeted drugs~\citep{Alexiou-2002, Maver-2009}, and in cancer treatments through magnetohyperthermia~\citep{Sato-2009, Verde-2012, Carriao-2016, Rego-2019, GOPIKA-2021, Moacua-2022}.
Improved targeted therapies, accurate diagnoses, and other innovative biomedical devices can
emerge by combining the magnetic response of ferrofluids with the tailored functionalization of nanoparticle surfaces, thereby opening new frontiers in the field of nanomedicine.

A major issue in nanotechnological biomedicine~\citep{Patil-2021} is the biocompatibility of the synthetic materials, being necessary to carry out extensive studies to ensure that they are pharmacologically inert for incorporation into living systems.
Ferrofluid-based biomedical applications impose strong constraints for practical use.
The particles must be chemically atoxic and the applied field's magnitude and frequency must not be unrestrictedly high to as to harm the body.
Also, the attraction between the nanoparticles must not cause extensive agglomeration. It is known that agglomeration might make body clearance difficult, making the substance remain in the body for an extended time. For drug-delivery applications, the concentration of nanoparticles might result in local deposition, overdosing the tissue and leading to cell death~\citep{bruinink2015effect,epple2018review}. For instance, it was found that the cytotoxicity of hydroxyapatite nanoparticles increases with agglomeration~\citep{andree2024effect}. Agglomeration-induced damages to DNA are reported for carbon nanotubes~\citep{wick2007degree} and anatase/rutile TiO$_2$ nanoparticles~\citep{magdolenova2012impact}. Therefore, an appropriate description of agglomeration is an important feature of colloidal theories.

Modifications to particle properties, such as surface functionalization or the addition of surfactants, can also alter the interaction energy between the particles and thus the colloidal stability of the ferrofluid.
The DLVO theory is the standard formalism to describe colloidal particle-particle interactions \citep{tufenkji2004deviation,Israelachvili-1992,hoek2003effect,PAULA-2007, Paula-2009,russel1991colloidal,shen2012application}. In its simplest form, it comprises two components \citep{liang2007interaction}: the van der Waals and the electric double-layer interactions. In general, several approaches are available to describe them. The van der Waals contribution can be modeled using either a macroscopic or a microscopic picture \citep{liang2007interaction}. For instance, the Lifshitz macroscopic theory treats the colloid as a continuum \citep{lifshitz1992theory}. Another alternative is to ground the formalism in quantum-mechanical principles, as done by London \citep{london1930theorie}. Concerning the DLVO, the Hamaker approach is used, according to which the interactions are calculated between simplified geometric shapes, such as planes and spheres \citep{Hamaker-1937,Israelachvili-1992}.

However, the DLVO theory has its limitations. In the small-separation regime, the van der Waals interaction can diverge if not properly controlled, leading to nonphysical trapping regions in theoretical simulations. When two nanoparticles come too close, they snap together and become inseparable, forming an unrealistically strong binding. Consequently, this limitation leads to a poor description of agglomeration and complex formation. The conventional workaround in is to implement a cutoff distance. Then, two neighboring nanoparticles closer than this limit have the same attractive energy. However, this approach also has its own flaws, as the threshold distance is an arbitrary parameter and may still enforce trapping regions. In this context, developing improved strategies is crucial for both theoretical and practical perspectives. A more detailed theory of colloids for small-separation regimes might deliver a better description of agglomeration mechanisms, potentially offering better insights into how to mitigate them.

To study different modeling approaches for short-separation interactions between nanoparticles, we rely on Monte Carlo simulations. Computational simulation methods, such as Monte Carlo and molecular dynamics have been used extensively to evaluate the effect of nanoparticle interactions on ferrofluid's properties~\citep{Ono-2015}, considering factors such as concentration~\citep{Castro-2006}, pH~\citep{Paula-2023}, ionic strength ~\citep{Brancolini-2022}, surface charge~\citep{Campos-2009}, particle size ~\citep{DENG-2016}, polymer architecture \citep{Mostarac-2022}, external magnetic field ~\citep{Meriguet-2004, Paula-2019}, and confinement (e.g. inside liposomes) ~\citep{Salvador-2016}.

This work was mainly focused on the influence of size dispersion and short-range interaction on the self-organization of nanoparticles in magnetic colloids. We compared the results of three interparticle interaction models, differing mostly in their short-distance descriptions, used in Monte Carlo simulations of this reference system. We took as our study case a system formed by a magnetic fluid based on tartrate-coated magnetite nanoparticles~\citep{Bakuzis-2013} that was synthesized by thermal coprecipitation and dispersed in water at $pH\sim7$ and a \ce{NaCl} concentration of \SI{0.15}{\molar} (physiological conditions).
The particle diameters were fitted into a log-normal distributed with modal diameter $D_0 = \SI{7.17}{\nano\meter}$ and dispersion $\sigma = \num{0.24}$ (for the sake of comparison, we simulated the system with and without diameter polydispersity).

In Section \ref{sec:models} we introduce the three models, proposing different solutions for correcting the small-distance interaction energy. In Section \ref{sec:simulation} we describe the simulation methods for studying the ferrofluid system, and proceed with an analysis of the energy extensivity properties of the different models, having in mind that the finite range interactions could lead to nonlinear scaling of energy with system size. We proceed with the presentation of our main results in Section \ref{sec:results} followed by a final discussion and conclusion.

\section{Models}
\label{sec:models}


Colloids are commonly modeled as an ensemble of $N$ interacting nanoparticles in a dispersion medium.
The balance between the particle interactions and eventual external agents dictates the macroscopic properties.
In our case, the physical description of a colloid involves summing the van der Waals ($U_{vdw}$), electric ($U_{edl}$), steric ($U_{ste}$), and magnetic interactions ($U_{mag}$).
\begin{equation}
    U = U_{vdw} + U_{edl} + U_{ste}  + U_{mag}.
\end{equation}

The three interaction models between the nanocolloidal particles differ in how the van der Waals and the electronic repulsion (EDL) interactions are described.
All other interactions pertaining to the description of the ferrofluid remain the same regardless of the model.

\begin{figure}
    \centering
    \includegraphics[width=0.8\linewidth]{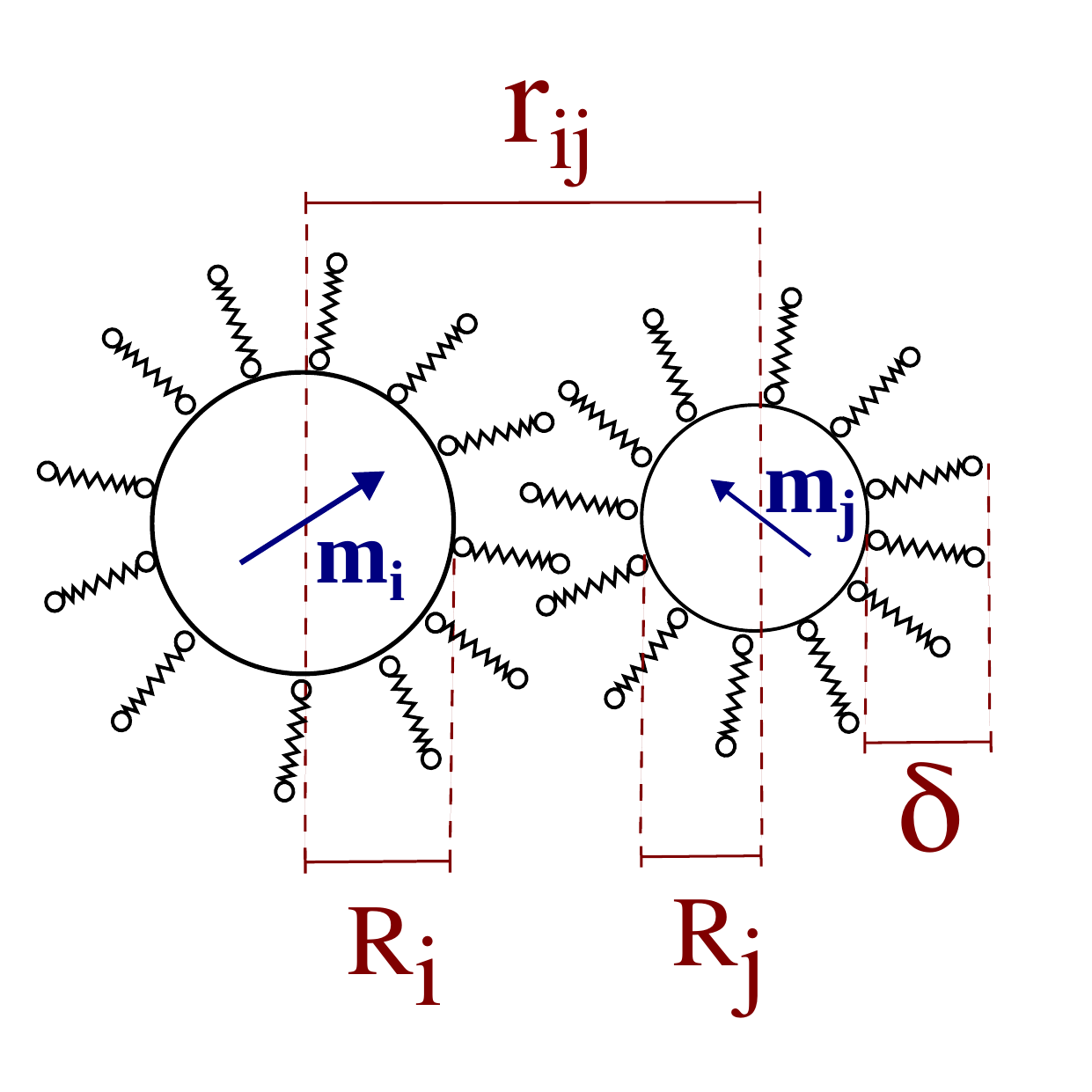}
    \caption{Ilustration of pair of colloidal magnetic nanoparticles coated with surfactant molecules. }
    \label{fig:particlepair}
\end{figure}

First, consider the interactions that are common to all models.
Let $R_i$ and $R_j$ be the radii of nanoparticles $i$ and $j$
positioned at $\vec{r}_i$ and $\vec{r}_j$, respectively.
Then, $r_{i,j} \equiv | \vec{r}_{i}-\vec{r}_{j}|$ is the corresponding
center-to-center distance as illustrated in Figure \ref{fig:particlepair}.
The magnetic moment of the i-th nanoparticle is
\begin{eqnarray}
    \vec{m}_i = M V_{mag(i)} \hat{u}_i,
\end{eqnarray}
where \(\hat{u}_i\) is the unit vector along the direction of the magnetic moment of the nanoparticle, \(M\) is the magnetization of the magnetite, and \(V_{mag(i)}\) is the ``magnetic volume'' of the nanoparticle, given by
\begin{eqnarray}
	V_{mag(i)} = \frac{4\pi}{3}(R_i-\delta_s)^3,
	\label{eq:Vmag}
\end{eqnarray}
in which $\delta_s$ is the length of its non-magnetic shell.
Therefore, the dipolar interaction energy between the nanoparticles is expressed as
 \begin{eqnarray}
	U_{mag}(r_{ij},\vec{m_{i}},\vec{m_{j}}) = \frac{\mu}{4\pi} \left[ \frac{\vec{m_{i}}\cdot \vec{m_{j}} - 3 (\vec{m_{i}}\cdot \hat{r}_{ij})(\vec{m_{j}}\cdot \hat{r}_{ij})}{r_{ij}^{3}} \right],
	\label{eq:Umag}
\end{eqnarray}
where \(\mu\) is the magnetic permeability of the solvent.

The steric repulsion energy reads~\citep{Rosensweig-1997}
\begin{equation}
    U_{ste}(r_{ij}) =
    \begin{cases}
        \frac{\pi \xi k_B T}{2} & (2R_i) (2R_j) \left[ 2  - \frac{(r_{ij} - 2 R_{ij})}{\delta} - \frac{r_{ij}}{\delta} \ln \bigg( \frac{2(R_{ij}+\delta)}{r_{ij}} \bigg) \right] \\
        & \text{for} \quad r_{ij} < 2 (R_{ij} + \delta), \\
         0 & \text{otherwise},
    \end{cases}
   \label{eq:Uster}
\end{equation}
in which $\xi$ is the grafting parameter (surface density of the adsorbed molecules), $\delta$ is the length of the surfactant layers, and $R_{ij}$ is the arithmetic mean of $R_i$ and $R_j$,
\begin{equation}
    R_{ij} = \frac{R_i + R_j}{2}.
\end{equation}

Let us now describe the interactions that differ across the models.
The first model avoids the van der Waals divergence by limiting the separation distance to a predefined threshold ($s_{min}$).
Any pairs whose surfaces are closer than this limit will have the same energy as if the distance is $s_{min}$.
The second model overcomes the divergence by considering the length of atomic bonds, effectively adjusting the separation distance in the energy expressions so that the divergence range is never reached.
The third model involves replacing $U_{vdw}$, for small separations, with the cohesive energy combined with the Born-Mayer repulsion.
We set $s_{min} = \SI{0.01}{\nano\meter}$ for the three models to compare the different approaches at small separations where the system transitions from a mesoscopic to a microscopic realm.
This value is close to the Born-Mayer decay parameter $L = \SI{0.03084}{\nano\meter}$,
while still above the order of magnitude in which finer effects such as atomic bond vibrations become significant \citep{Konaka-1970,Scheele-2024}.

\begin{figure*}[ht]
  \begin{center}
    \begin{subfigure}{0.33\hsize}
	\centering
	\includegraphics[width=\linewidth]{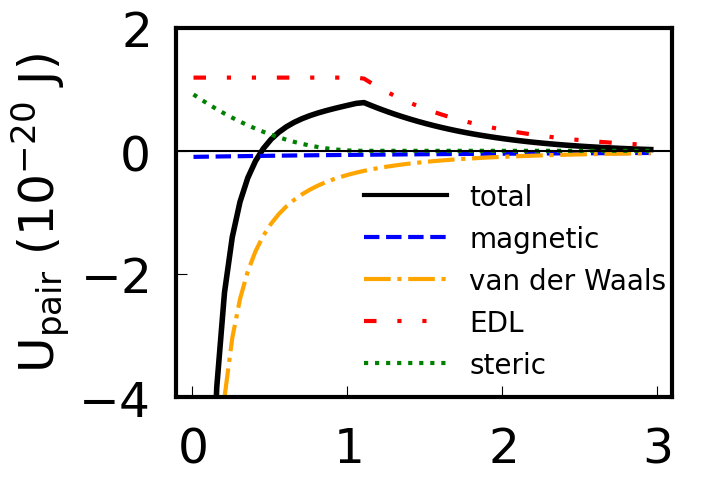} \\
        \includegraphics[width=\linewidth]{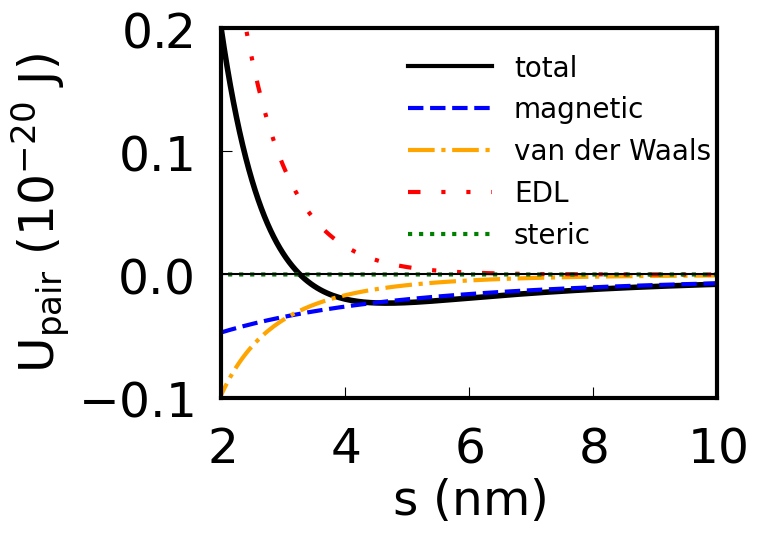}
	\caption{Model 1}
	\label{fig:u_vs_s_model_1}
    \end{subfigure}
    \begin{subfigure}{0.3\hsize}
	\centering
	\includegraphics[width=\linewidth]{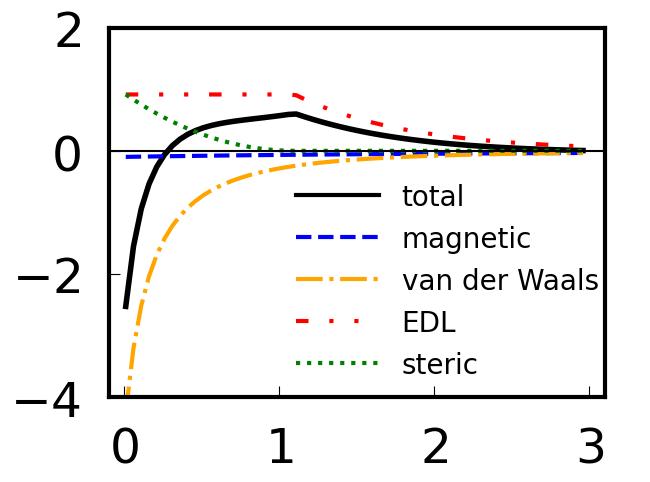} \\
  	\includegraphics[width=\linewidth]{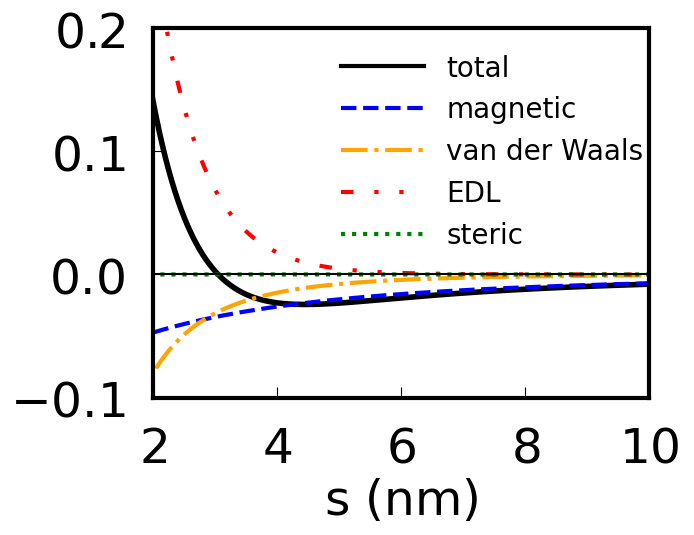}
        \caption{Model 2}
	\label{fig:u_vs_s_model_2}
    \end{subfigure}
    \begin{subfigure}{0.3\hsize}
	\centering
	\includegraphics[width=\linewidth]{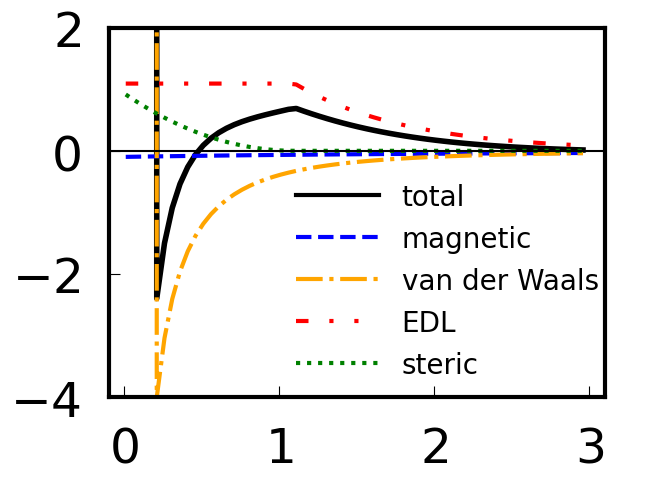}
        \includegraphics[width=\linewidth]{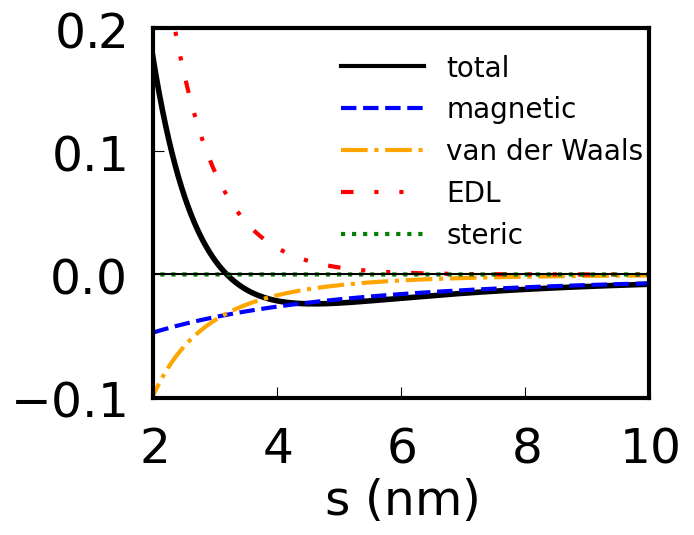}
        \caption{Model 3}
        \label{fig:u_vs_s_model_3}
    \end{subfigure}
  \end{center}
  \caption{Total energy of a nanoparticle pair ($U_{pair}$), according to each model, as a function of the distance between their surfaces ($s \equiv s_{ij} = r_{ij}-2R_{ij}$). The two particles are identical, with radii $R_i = R_j = \SI{3.585}{\nano\meter}$ ($= D_0/2$), disposed with magnetic dipoles aligned in a head-to-tail fashion (note the negative sign of the magnetic contribution). Other parameters are set as given in \ref{sec:appendix:parameters}. The graphs in the top row show small distances (and consequently high energy values). In contrast, the graphs in the bottom row show the corresponding curves at larger distances (and low energy), evidencing the long-range nature of magnetic interactions. Models 2 and 3 have different strategies to deal with the divergence of the van der Waals energy, while it is barely treated in Model 1.}
  \label{fig:u_vs_s}
\end{figure*}

\subsection{Model 1}

In the first model, the van der Waals energy value decreases as the surface-surface distance moves towards its minimum cutoff value of $s_{min} = \SI{0.01}{\nano\meter}$, and then it is held constant for smaller distances ($s < s_{min}$).

The van der Waals energy between two particles $R_i$ and $R_j$ is written as~\citep{Hamaker-1937}
    \begin{eqnarray}
		U_{vdw}(r_{ij}) = -\frac{A}{6}\left[ \frac{2R_iR_j}{r_{ij}^{2}-(R_i+R_j)^2}
            + \frac{2R_iR_j}{r_{ij}^{2}-(R_i-R_j)^2} \right. \\ \nonumber
            \left.
            + ln \left( \frac{r_{ij}^{2}-(R_i+R_j)^2}{r_{ij}^{2}-(R_i-R_j)^2} \right) \right].
		\label{eq:Uvdw1}
    \end{eqnarray}
The value of \(s_{min}\) was set as \(\SI{0.01}{\nano\meter}\), leading to a deep well in the van der Waals energy, as shown in Figure~\ref{fig:u_vs_s_model_1}. The chosen $s_{min}$ is significantly lower than the conventional use of van der Waals interactions. We recall that the cutoff distance is a form of mitigating the energy divergence for small-range separations. Therefore, the value of $s_{min}$ is, within a physical range, arbitrary and often tuned to avoid sampling problems. Here, we opted for a lower $s_{min}$ to showcase the effect of the corrections proposed in the work via Models 2 and 3. 

The electrostatic repulsion is described using the linear superposition model~\citep{Russel-1989}, adapted to the fact that the surface charge is at the surfactant layer extremity:
    \begin{eqnarray}
	U_{edl}(r_{ij}) =  4 \pi k_B T (2\rho_{ion}) \lambda_D
              \left[ R_i^{\star} \Psi_{(R_i^{\star})} \right]
              \left[ R_j^{\star} \Psi_{(R_j^{\star})} \right] \\ \nonumber
            \times \exp \left(- \frac{r_{ij}-R_i^{\star}-R_j^{\star}}{\lambda_D} \right)  / r_{ij} ,
		\label{eq:Uedl1}
    \end{eqnarray}
where the function $\Psi_{(R)}$ is expressed as
    \begin{eqnarray}
        \Psi_{(R)} = 4  \tanh{\left(\frac{e\psi_{(R)}}{4k_BT}\right)},
    \end{eqnarray}
(from the two-flat-plane model via the Derjaguin approximation for two spheres) ensuring that $U_{edl}$ closely approximates the most accurate models for both small and large separations, and
    \begin{eqnarray}
	R^{\star} = R + \delta
    \end{eqnarray}
is the effective nanoparticle radius adjusted by $\delta$, the thickness of the surfactant layer (the length of the molecules), $k_B$ is the Boltzmann constant, $T$ is the absolute temperature, $e$ is the elementary charge, $\psi_{(R)}$ is the value of the electric potential on the surface of a nanoparticle of radius $R$ ($R_i$ or $R_j$ in Eq.~\ref{eq:Uedl1}), $s_{\sigma}$ is the distance between the charged surfaces, and $\lambda_D$ is the Debye length, given by
\begin{eqnarray}
	\lambda_D = \left(\frac{\epsilon k_B T}{(2\rho_{ion})z_{ion}^2e^2}\right)^{1/2},
	\label{eq:lambdaD}
\end{eqnarray}
where $\epsilon$ is the electrical permittivity of the solvent,  $\rho_{ion}$ is the ion concentration of the solvent, and $z_{ion}$ is the valence of the ions in the solution.
The system simulated in this study has a 1:1 electrolyte (\ce{NaCl}), therefore we used $z_{ion}=1$.

The value of $\psi_{(R)}$ is calculated as \citep{Hunter-1981}
\begin{eqnarray}
    \psi_{(R)} = \frac{k_B T}{z e} \operatorname{acosh}\left( \frac{\sigma_{(R)}^2}{4 \epsilon \rho_{ion} k_B T} \right),
    \label{eq:psii}
\end{eqnarray}
where $\sigma_{(R)}$ is the surface charge density of a nanoparticle of radius $R$, calculated through the relation~\citep{Bakuzis-2013}
\begin{eqnarray}
    \sigma_{(R)} = \left[ z e \left( \frac{R}{R+\delta} \right)^2 \right] \xi,
    \label{eq:sigmai}
\end{eqnarray}
which expresses the linear relation between the surface charge density and the surfactant density (\(\xi\)), considering that each surfactant molecule contributes with a charge \(ze\) at its extremity, located at a distance \(\delta\) from the nanoparticle surface. Note that $\psi_{(R)}$ carries a dependence on particle size via the surface charge density, consistent with the expected behavior of the electric potential under changes in particle size.

Note that the accuracy of the Derjaguin approximation is higher for small separation-to-diameter ratios. However, for larger distances, the interaction becomes progressively smaller, lowering its importance in the overall energy balance. Therefore, for the sake of simplicity, the same expression is applied for other separation distances since its importance is reduced. This represents the standard approach in simulating colloids within the DLVO formalism. Specifically, the EDL interaction, as present here, is the usual approach to model charge surface interactions in colloid models, extensively applied in the past to represent interactions in the same class of colloids \citep{kumar2010state,salvador2016characterization}. 
Model 1, described in this subsection, will be a reference for comparison with two other strategies, Models 2 and 3, described next. Importantly, Model 3 will specifically address the limitation in the description of the EDL interaction by introducing a new expression, derived via the Schnitzer–Morozov technique \citep{Schnitzer-2015}, which extends the validity of the Derjaguin approximation into the long-range regime.

\subsection{Model 2}

The second model considers the length of a typical atomic bond in the nanoparticle bulk ($L_B$) and prevents the divergence of the van der Waals energy by adding to the intersurface distance of the colloidal particle pairs, which is equivalent to changing the center-to-center distance as in
\begin{equation}
    r_{ij} \rightarrow r_{ij} + L_B.
    \label{eq:rijshift}
\end{equation}
Incorporating this distance into the center-to-center separation aims to preserve the qualitative characteristics of the interaction curve, with noticeable deviations emerging only at distances near $L_B$, where the divergence occurs (Figure \ref{fig:u_vs_s_model_2}).
This procedure is similar to a modification tested by White~\citep{White-2010} for the van der Waals interaction between planes.
We have chosen the value of a typical atomic bond distance of the maghemite bulk, $L_B = \SI{0.19}{\nano\meter}$, estimated for a \ce{Fe-O} bond \citep{Fdez-2016, Coduri-2020}.

Thus, the van der Waals energy between the nanoparticles $i$ and $j$ is then written as
    \begin{eqnarray}
            U_{vdw}(r_{ij})
            = - \frac{A}{6}\left[ \frac{2R_iR_j}{(r_{ij}+L_B)^{2}-(R_i+R_j)^2}\right.\\
            \nonumber
            \left.
            + \frac{2R_iR_j}{(r_{ij}+L_B)^{2}-(R_i-R_j)^2} 
            + ln \left( \frac{(r_{ij}+L_B)^{2}-(R_i+R_j)^2}{(r_{ij}+L_B)^{2}-(R_i-R_j)^2} \right) \right].
		\label{eq:Uvdw2}
    \end{eqnarray}
    
For the sake of consistency, the same modification is applied to the EDL interaction:
    \begin{eqnarray}
	U_{edl}(r_{ij}) = &4 \pi k_B T (2\rho_{ion}) \lambda_D \nonumber \\
            &\times \frac{\displaystyle\left[ 4 R_i^{\star} \tanh{\left(\frac{e\psi_{(R_i)}}{4k_BT}\right)} \right]
                  \left[ 4 R_j^{\star} \tanh{\left(\frac{e\psi_{(R_j)}}{4k_BT}\right)} \right]
                 }{\displaystyle r_{ij}+L_B} \\ \nonumber
            &\times \exp \left(- \frac{\displaystyle r_{ij}+L_B-R_i^{\star}-R_j^{\star}}{\displaystyle\lambda_D} \right).
		\label{eq:Uedl2}
    \end{eqnarray}
The values of $\psi_{(R_i)}$ (and $\psi_{(R_j)}$) are given by Eq.~\ref{eq:psii}, as in Model 1.

\subsection{Model 3}

Another approach consists of replacing, for small separations, the van der Waals energy with the Born-Mayer repulsion and the cohesive energy~\citep{Castro-2022}, so that the expression for the joint expression looks like
\begin{equation}
    U_{vdw-bm}(r_{ij}) 
    =\begin{cases}
		U_{vdw}(r_{ij})	 \hfill \text{ if } r_{ij} > R_{ij} + L_{B} ,  \\
		U_{vdw}(L_B)+U_{coh}(r_{ij})+U_{bm}(r_{ij}) \hfill \text{ otherwise} ,
    \end{cases}
\end{equation}
where $U_{vdw}(L_B)$ is the value of $U_{vdw}(r_{ij})$ for $r_{ij}=L_B$ (value shown in \ref{sec:appendix:parameters}), $U_{bm}$ is the energy related to the Born-Mayer repulsion~\citep{Lewis-1985}, given by
    \begin{eqnarray}
	U_{bm} = A_{bm} \exp\left(-\frac{r_{ij}-R_{ij}}{L_{bm}}\right),
    \end{eqnarray}
and
    \begin{eqnarray}
	U_{coh} = A_{int} W_{coh},
    \end{eqnarray}	
where $A_{int}$ is the area between the spheres in which their surfaces are close enough to form atomic bonds (details of the calculation were shown in a previous article \citep{Castro-2022}) and $W_{coh}$ is the cohesion work (value shown in \ref{sec:appendix:parameters}).

The authors also propose a change in the electrostatic interaction energy that makes it accurate at all distances, using a technique introduced by Schnitzer and Morozov~\citep{Schnitzer-2015}.
The resulting expression is
\begin{eqnarray}
	U_{edl-sm(r_{ij})} = - 4 \pi k_B T (2\rho_{ion}) \lambda_D
		\left[ 4 R_i^{\star} \tanh{\left(\frac{e\psi_i}{4k_BT}\right)} \right]\\
            \left[ 4 R_j^{\star} \tanh{\left(\frac{e\psi_j}{4k_BT}\right)} \right] \nonumber 
		\times e^{(R_i+R_j)/\lambda_D} Ei\left(-\frac{r_{ij}-2\delta}{\lambda_D}\right),
	\label{eq:Uedl3}
\end{eqnarray}
where $\rho_{ion}$ is the concentration of ions in the solvent, $\psi_i$ and $\psi_j$ are the electrical potential values on the surfaces of the spheres, $Ei$ is the exponential integral function, and $\lambda_D$ is the Debye length.

The energy curves of Model 3 are depicted in Figure~\ref{fig:u_vs_s_model_3}.
Although the energy curves provided by the three models are very similar for long distances (bottom row of Figure~\ref{fig:u_vs_s}), they exhibit significant differences for short distances (top row) in terms of shape and minimum value.
Therefore, the models represent the colloid using distinct physical descriptions that we shall compare in the results of this study.

\section{Simulation}
\label{sec:simulation}

The system was studied using Monte Carlo simulations, through an adaptation of the Metropolis-Hastings algorithm. Each simulation starts with a random configuration and small changes in the particle positions and orientations are made in successive steps.
A newly generated configuration is accepted whenever the system energy decreases.
If the energy increases by $\Delta{E}$, the new configuration is accepted with a probability proportional to $\exp(-\beta\Delta{E})$, with $\beta=(k_BT)^{-1}$.
The scale of such changes is dynamically adjusted such that 50$\%$ of the generated configurations are accepted, by following the algorithm rules.
After every \num{50000} steps, the program calculates how much the ever-reached minimum value of total energy has decreased: sampling begins when this variation is less than \num{0.05}, continues for at least other \num{50000} steps, and finishes when the variation is less than \num{0.005}.
Those parameters are summarized in \ref{sec:appendix:parameters}.
The quantities of interest are averaged over the sampling steps of at least ten independent simulations with each set of physical parameters.

To represent a sample from a macroscopic object, we simulated the colloids using periodic boundary conditions with a minimum image. The pair interactions between two particles are calculated over the smallest distance between them, considering their images through the periodic boundaries. Also, the periodic boundary conditions are used at the Monte Carlo steps: When generating a new sample configuration, the nanoparticles' positions are weakly shifted. If the shift moves the particle crosses the boundaries of the original box, it is reinserted near the opposite face of the box, according to the periodic conditions.

Initially, we assess the total energy extensivity of the three models. Given the finite range of pair interactions, we can expect some deviation from a linear scaling of energy with system size (increasing the number of particles and the volume of the system while preserving the volume fraction of particles). It is easy to see that if we split a colloidal system with a plane separating it into parts A and B, the total energy $U$ will not be just the sum of the energies of the separate parts, $U_{A}+U_{B}$. There can be a significant contribution to the total energy that is associated with pairs of particles near the separating plane, with each particle of the pair on a different side. To evaluate this contribution, we simulated the colloidal solution with increasingly higher particle numbers and evaluated their corresponding energies. The particle concentration was kept constant at \SI{5}{\percent} (volume fraction $\phi$), a relatively high value that forces more particles into small separations. The interactions between the nanoparticles are higher in this regime, allowing a better evaluation of the system deviation from extensivity. After that, we simulated the ferrofluid at a lower concentration of $\phi = \SI{0.47}{\percent}$, which was investigated in detail in our experimental reference \citep{Bakuzis-2013}.
The simulations included some variations as a means of comparing the effects of the characteristics and conditions of the colloid: monodisperse and polydisperse diameter distributions; non-magnetic and magnetic particles; and zero and finite applied magnetic fields.
The monodisperse system corresponds to a log-normal distribution with a dispersion parameter of zero ($\sigma_R = 0$), meaning that all diameters are equal to the median diameter of the polydisperse system. All results presented in the next section are based on averages took over 10 separate Monte Carlo simulations, to account for the random aspect of these simulations.
Other parameters are shown in \ref{sec:appendix:parameters}.

\section{Results}
\label{sec:results}

\textbf{Dispersity differentially affects colloidal structure depending on short-range interaction models.} We start characterizing the magnetic colloids under the three models by looking into the average first neighbor distance $\langle s_{1} \rangle$ for a volume fraction $ \phi = 5\% $ (see Table \ref{tab:s1st-highconc}; the dependence of $\langle s_{1} \rangle$ with the number of particles is shown in the Supplementary Figure S1).

Model 1 and Model 2 lead to first neighbor distances that are larger polydisperse than in the monodisperse scenario.
This is due to the EDL repulsion increasing faster than the magnetic dipolar and van der Waals attractions, for increasing diameters in the short range.
The log-normal distribution is skewed towards larger diameters, which favors higher barriers that prevent nanoparticles from reaching the primary energy minima (see Supplementary Figure S2).

Model 3 resulted in more sparsely distributed nanoparticles, a consequence of the short-range repulsions.
However, in contrast with the other models, the values of $\langle s_{1} \rangle$ are now larger for the monodisperse system.
This can be understood by looking into the position of the secondary energy minima, which move to shorter distances and become deeper due to the increased magnetic attraction of larger particles (see Supplementary Figure S3). This moving to shorter distances and deepening of the secondary energy minimum is still observed for pairs of particles with different sizes, as one increases the radius of either particle in a pair (Supplementary Figure S4).

\begin{table}
    \centering
    \caption{Average first neighbor distance $\langle s_{1} \rangle$ for monodisperse and polydisperse systems, at a volume fraction of $5\%$ simulated through different models, with standard errors. The short-range repulsions of Model 3 caused overall higher values, especially for the monodisperse system (in contrast with the other models, with higher distances for the polydisperse system).}
    \label{tab:s1st-highconc}
		
    \begin{tabular}{|>{\centering\arraybackslash}m{1.5cm}|>{\centering\arraybackslash}m{3.0cm}|>{\centering\arraybackslash}m{3.0cm}|}
        \hline
        \multirow{2}{*}{Model} & \multicolumn{2}{c|}{Average first neighbor distance (\si{\nano\meter})} \\
        \cline{2-3}
            &	monodisperse	& polydisperse \\ \hline
            1			&	\num{2.17(4)}	&	\num{2.61(4)} \\ 
            2			&	\num{2.06(5)}	&	\num{2.57(5)} \\
            3			&	\num{3.12(6)}	&	\num{2.74(4)} \\
        \hline
    \end{tabular}
\end{table}

We also analyzed $\langle s_{1} \rangle$ for the ferrofluid at the lower volume fraction of $0.47\%$, which characterizes the real sample \citep{Bakuzis-2013}. We considered both non-magnetic and magnetic colloids, with and without an external magnetic field (Table \ref{tab:s1st-mono} for monodisperse and Table \ref{tab:s1st-poli} for polidisperse scenarios).
For this decreased volume fraction, the average first-neighbor distance is larger (compare to Table \ref{tab:s1st-highconc}), and small-range repulsions included in Model 3 are only relevant for a minority of the nanoparticles.
The differences between the $\langle s_{1} \rangle$ values for non-magnetic and magnetic colloids at zero field, and magnetic colloids at \SI{1}{\tesla} are not significant, considering the uncertainties.
Our focus is on the nanoparticle interactions, more easily studied in the magnetic colloid at zero field.

We also compare Models 2 and 3 by looking into how they change the system compared to Model 1, which is an established formalism in colloid simulations. Both Models 2 and 3 predict equilibrium distances smaller than those predicted by Model 1 (Supplementary Figure S5). For pairs of particles with fixed raii sum $R_i + R_j$, the distinctions between Model 1 and both Models 2 and 3 in more pronounced in the monodisperse scenario ($R_{i} = R_{j}$). This is true for both the position and the energy level of the secondary equilibrium. The difference between Models 1 and 2 in the position of the secondary equilibrium presents a higher variability over particles' radii combinations, while Model 3 differs from Model 1 through a more systematic and better-conserved shift. Compared to Model 1, both Models 1 and 3 lead to increased energy depths, indicating an improved description of agglomeration mechanisms. The stronger secondary minimum may serve as a competing factor in the formation of strongly bound nanoparticles that were artificially trapped in the first minimum in Model 1.

\begin{table}
    \centering
    \caption{Average first neighbor distance $\langle s_{1} \rangle$ for the monodisperse systems simulated through different models, with standard errors,  at a volume fraction of $0.47\%$ \citep{Bakuzis-2013}.
    The values given by Model 3 are slightly higher, with a more significant disagreement with the other models, considering the standard errors.}
    \label{tab:s1st-mono}
		
    \begin{tabular}{|>{\centering\arraybackslash}m{0.8cm}|>{\centering\arraybackslash}m{2.0cm}|>{\centering\arraybackslash}m{2.0cm}|>{\centering\arraybackslash}m{2.0cm}|}
        \hline
        \multirow{2}{*}{Model} & \multicolumn{3}{c|}{Average first neighbor distance (\si{\nano\meter})} \\
        \cline{2-4}
            &	non-magn.	&	magnetic $B = \SI{0}{\tesla}$	& magnetic $B = \SI{1}{\tesla}$ \\
        \hline
            1			&	\num{11.88(7)}	&	\num{12.24(98)}	&	\num{12.11(9)} \\ 
            2			&	\num{11.80(8)}	&	\num{11.94(17)}	&	\num{12.24(92)} \\
            3			&	\num{12.45(55)}	&	\num{12.37(75)}	&	\num{12.34(87)} \\
        \hline
    \end{tabular}

\end{table}

\begin{table}
    \centering
    \caption{Average first neighbor distance $\langle s_{1} \rangle$ for the polydisperse system simulated through different models, with standard errors. The differences between the values are negligible, except for the slightly higher values of Model 3 for the non-magnetic system. This is explained by the larger particles that increase the EDL repulsion, which, in magnetic colloids, is counterbalanced by the increase of the magnetic dipolar attraction.}
    \label{tab:s1st-poli}
		
    \begin{tabular}{|>{\centering\arraybackslash}m{0.8cm}|>{\centering\arraybackslash}m{2.0cm}|>{\centering\arraybackslash}m{2.0cm}|>{\centering\arraybackslash}m{2.0cm}|}
        \hline
        \multirow{2}{*}{Model} & \multicolumn{3}{c|}{Average first neighbor distance (\si{\nano\meter})} \\
        \cline{2-4}
            &	non-magn.	&	magnetic $B = \SI{0}{\tesla}$	& magnetic  $B = \SI{1}{\tesla}$ \\ \hline
            1			&	\num{13.31(12)}	&	\num{13.47(12)}	&	\num{13.42(11)} \\ 
            2			&	\num{13.50(13)}	&	\num{13.20(13)}	&	\num{13.44(11)} \\
            3			&	\num{13.61(9)}	&	\num{13.76(13)}	&	\num{13.75(9)} \\
        \hline
    \end{tabular}
\end{table}

\begin{figure}[ht]
\begin{center}
\begin{subfigure}{0.8\columnwidth}
\centering
\includegraphics[width=\linewidth]{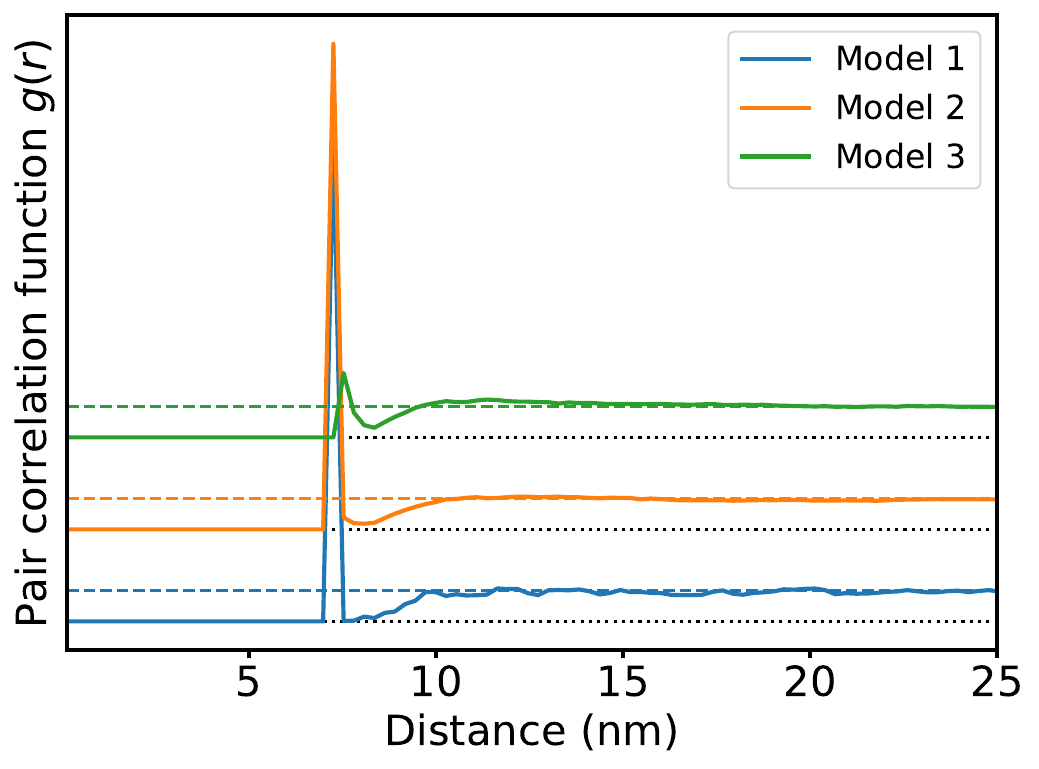}
\caption{Monodisperse}
\label{fig:gr_mono}
\end{subfigure} \\
\begin{subfigure}{0.8\columnwidth}
\centering
\includegraphics[width=\linewidth]{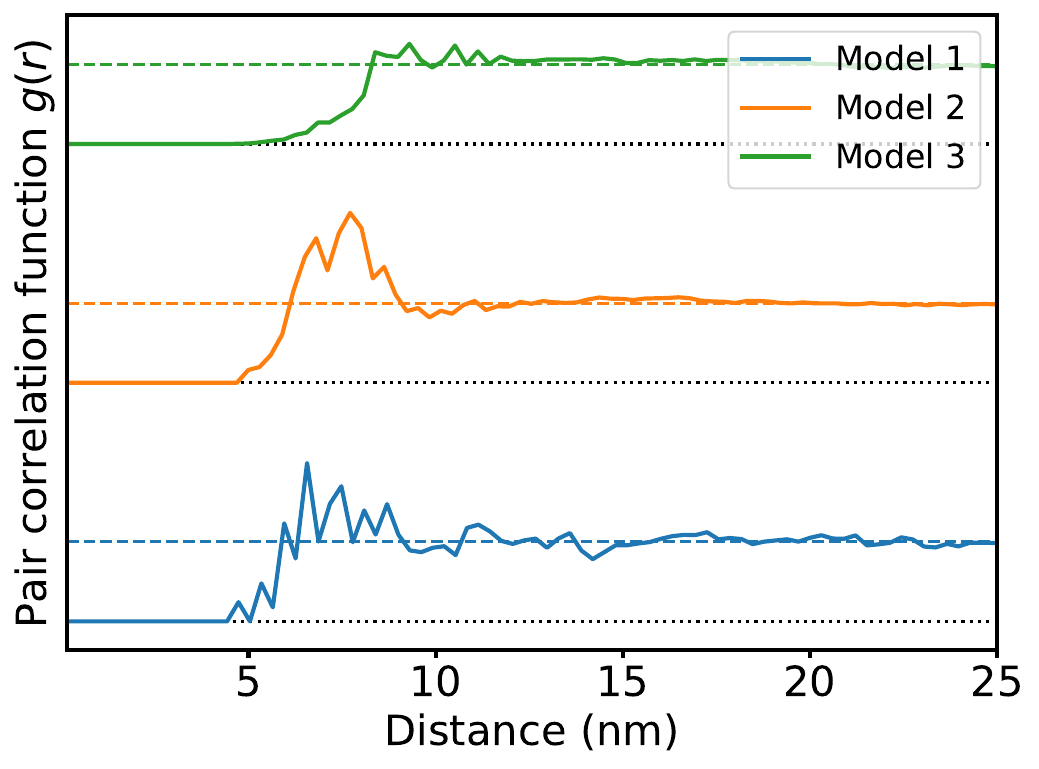}
\caption{Polydisperse}
\label{fig:gr_poli}
\end{subfigure}
\caption{Pair correlation function for the mono and polydisperse systems in the zero field condition, simulated through the three models, averaged over 10 simulations. Higher peaks are observed for Models 1 and 2, and the lowest ones with Model 3. Diameter polydispersity makes the peaks more dispersed. The curves in different colors are shifted vertically for easier comparison, with the dotted lines corresponding to the 0 particle baseline, and the colored dashed curves corresponding to the large distance limit $g(r) \rightarrow 1$.}
\label{fig:gr_r}
\end{center}
\end{figure}

\textbf{Models 2 and 3 improve Monte Carlo sampling of the system.} Going into more detail of the colloidal structure, we look into the pair correlation function $g(r)$ (Figures \ref{fig:gr_mono} for mono and \ref{fig:gr_poli} polydisperse systems in the zero field condition).
As expected, the peak heights are proportional to the energy well depths (Figure \ref{fig:u_vs_s}), with Models 1 and 2 resulting in higher peaks and Model 3 in lower ones.
The polydisperse systems pair-correlation functions exhibit more dispersed peaks, as a result of the statistical variability of the nanoparticle diameters, making multiple separation distances available depending on the radii of particles in a pair. However, it is important to note that the number and position of these multiple peaks can be sensitive to fluctuations in the Monte Carlo simulations.   

\begin{figure*}
\begin{centering}
\begin{tabular}{cc}
  \begin{tabular}{ccc}
    \begin{subfigure}{0.23\hsize}
	\centering
	\includegraphics[width=\linewidth]{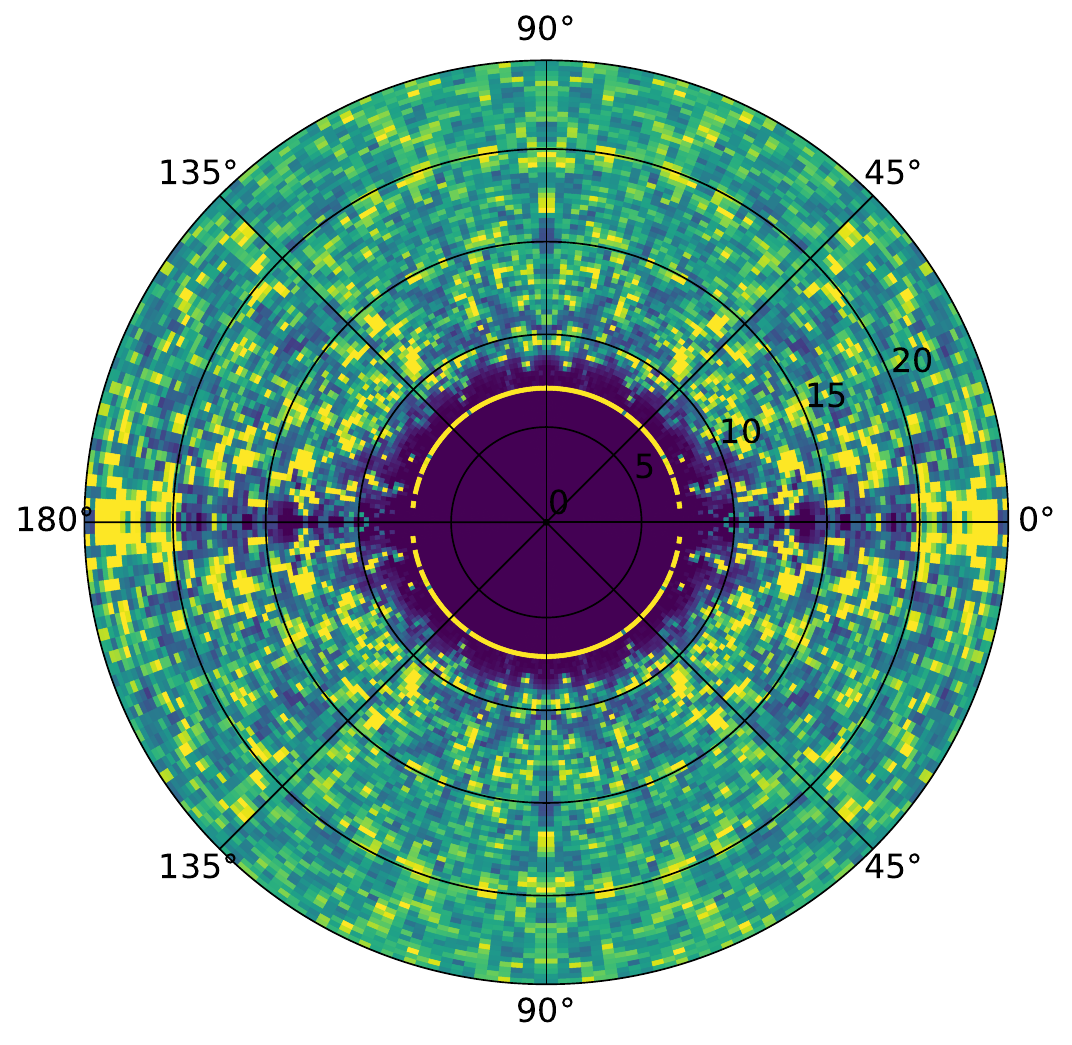}
	\caption{Model 1, monodisperse}
	\label{fig:gr_rt_model_1_monod}
    \end{subfigure}
    &
    \begin{subfigure}{0.23\hsize}
	\centering
	\includegraphics[width=\linewidth]{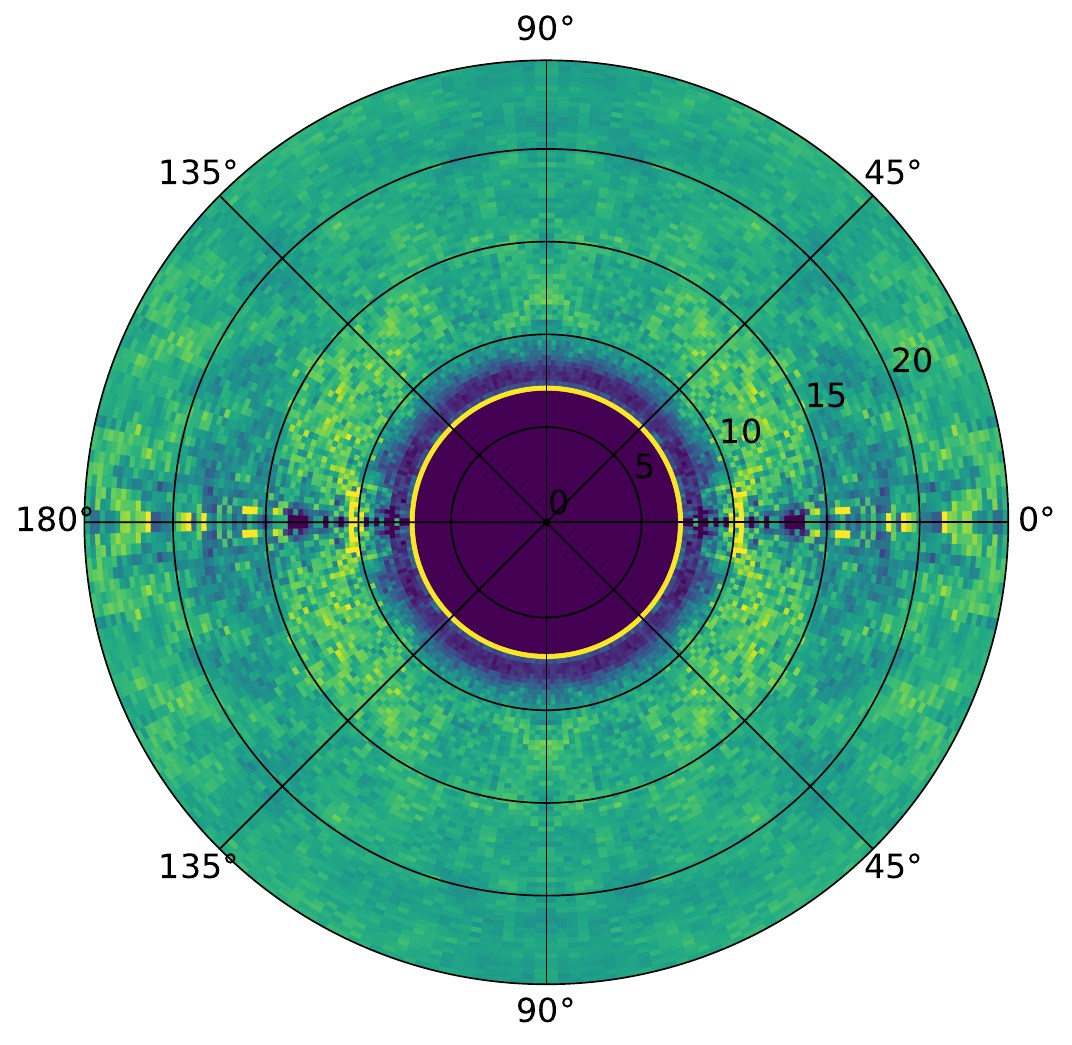}
	\caption{Model 2, monodisperse}
	\label{fig:gr_rt_model_2_monod}
    \end{subfigure}
    &
    \begin{subfigure}{0.23\hsize}
	\centering
	\includegraphics[width=\linewidth]{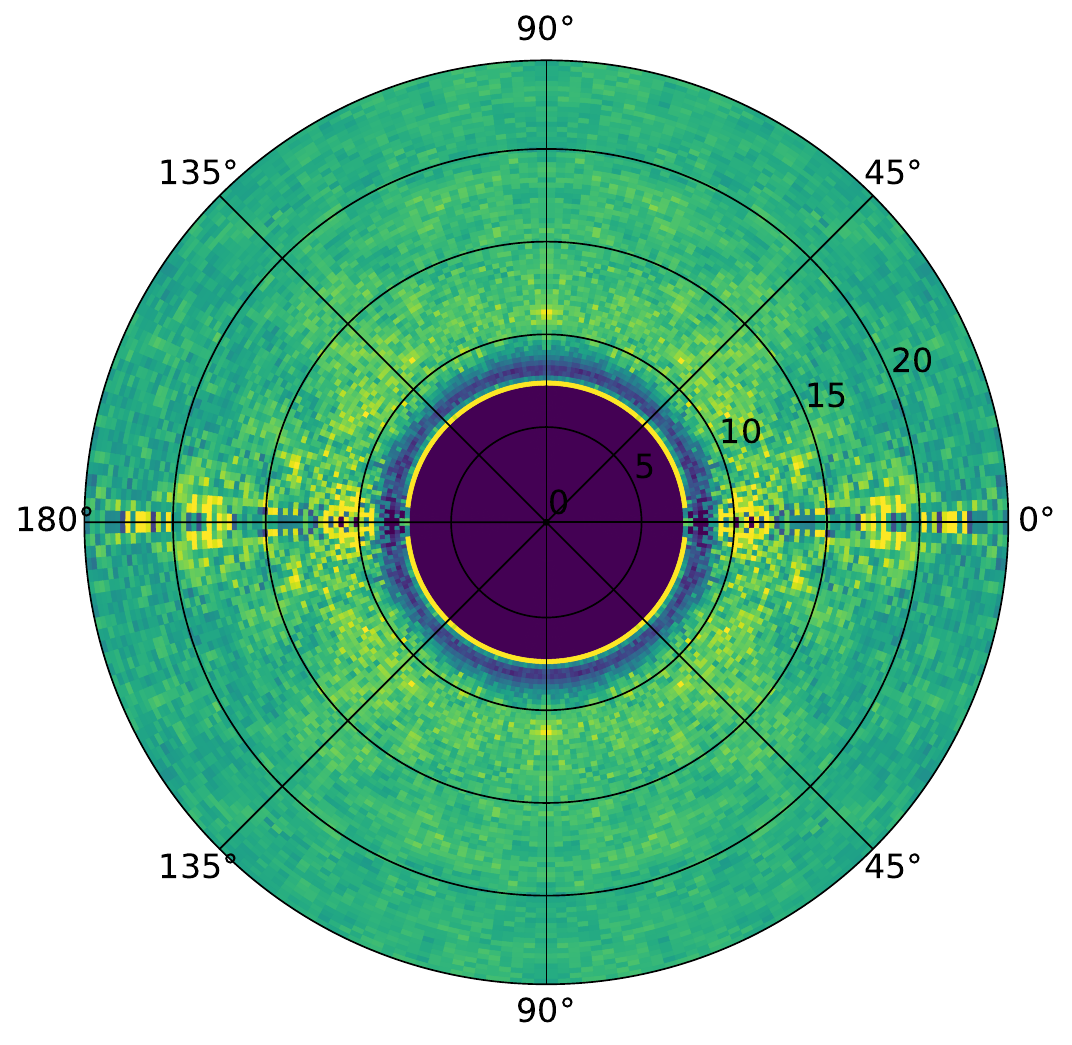}
	\caption{Model 3, monodisperse}
	\label{fig:gr_rt_model_3_monod}
    \end{subfigure}
    \\
    \begin{subfigure}{0.23\hsize}
	\centering
	\includegraphics[width=\linewidth]{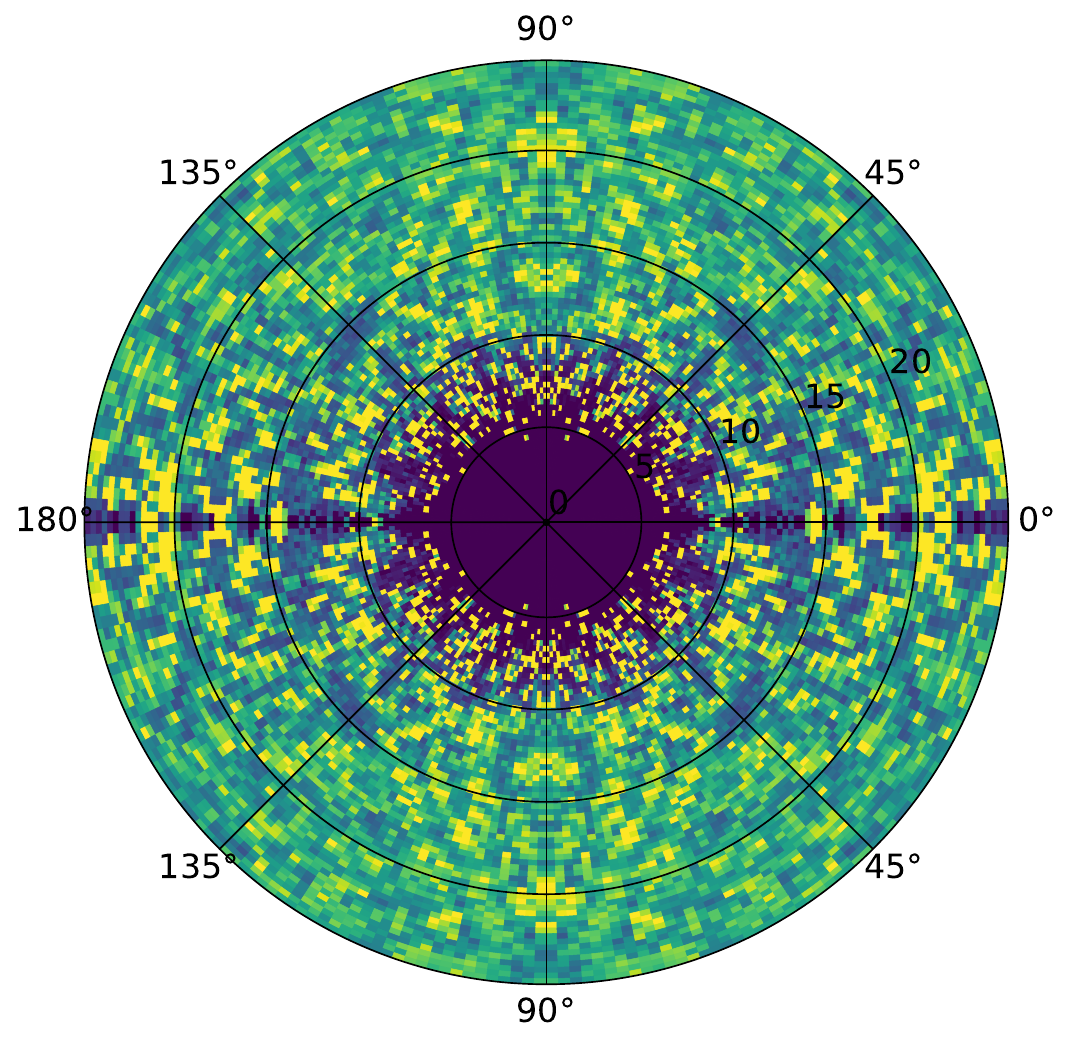}
	\caption{Model 1, polydisperse}
	\label{fig:gr_rt_model_1_polid}
    \end{subfigure}
    &
    \begin{subfigure}{0.23\hsize}
	\centering
	\includegraphics[width=\linewidth]{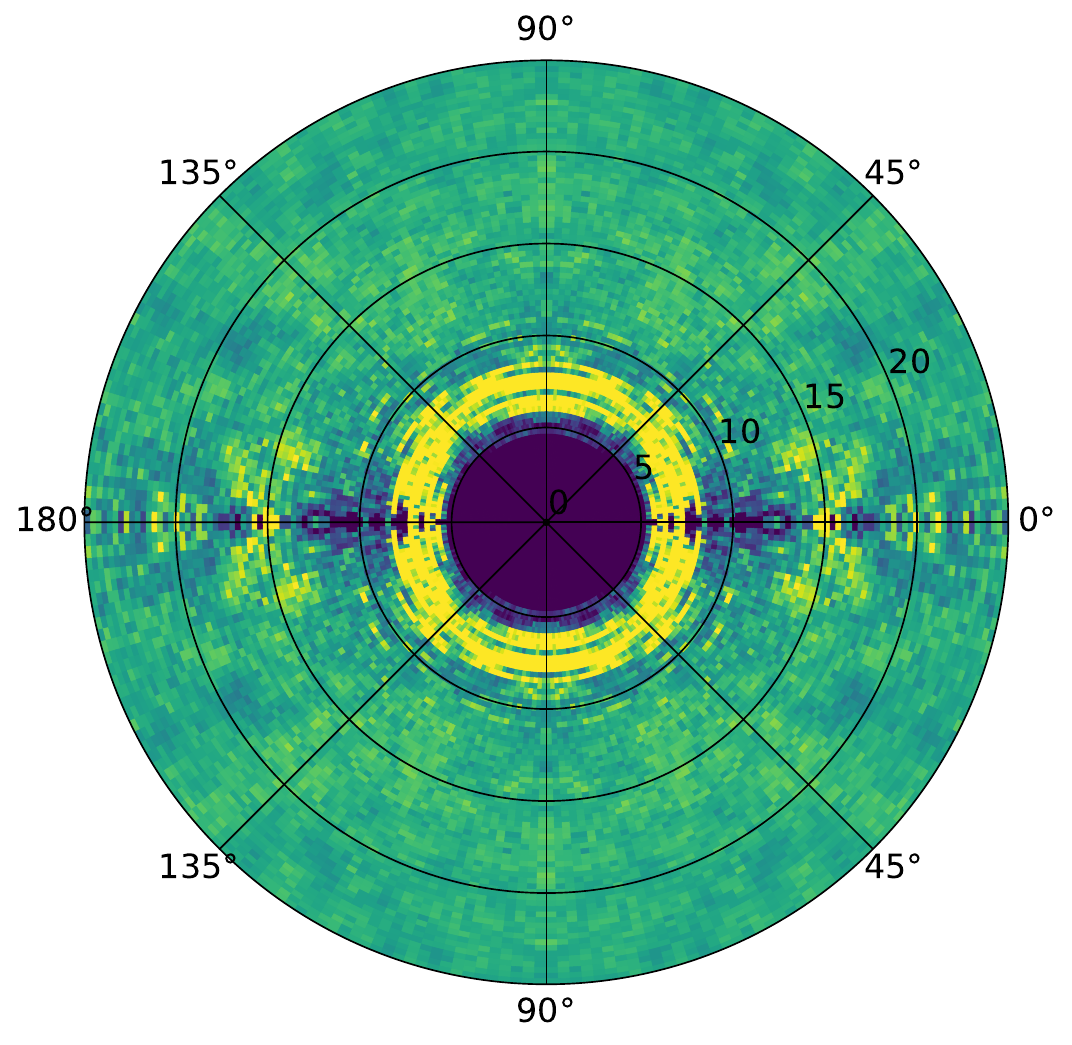}
	\caption{Model 2, polydisperse}
	\label{fig:gr_rt_model_2_polid}
    \end{subfigure}
    &
    \begin{subfigure}{0.23\hsize}
	\centering
	\includegraphics[width=\linewidth]{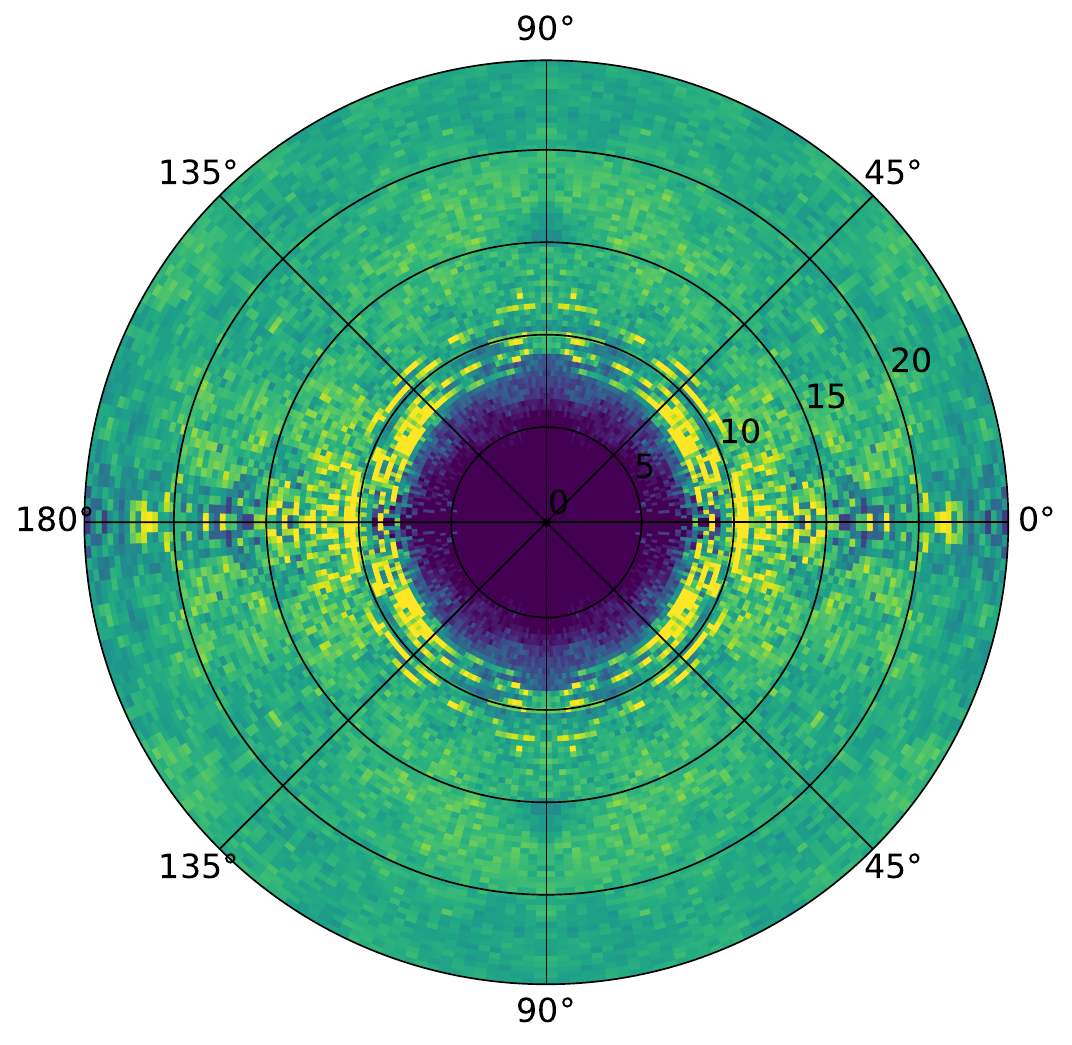}
	\caption{Model 3, polydisperse}
	\label{fig:gr_rt_model_3_polid}
    \end{subfigure}
    \\
  \end{tabular}
  &
  \begin{tabular}{c}
    \includegraphics[width=0.07\linewidth]{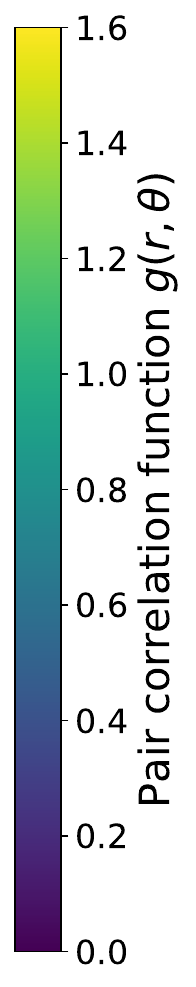}
  \end{tabular}
\end{tabular}
    \caption{\textbf{Models 2 and 3 improve Monte Carlo sampling of the system.} Pair correlation function $g(r,\theta)$, simulated through the three models, for magnetic systems at zero applied field condition, averaged over 10 simulations. Volume fraction of $\phi=0.47\%$. $r$ is in units of nanometers. The horizontal axis ($\theta=0\degree$) in the heatmaps coincides with the $z$-axis of a coordinate system $xyz$. The spots observed for Model 1 are caused by particle pairs trapped in local minima or the deep well as two surfaces approach each other. The dispersal of the first-neighbor-distance ring in the bottom row highlights the influence of polydispersity on the magnetic structure. For improved visualization, the colormap is limited to 1.6, and all points with $g(r,\theta)>1.6$ are shown in the same color.}
    \label{fig:gr_rt}
\end{centering}
\end{figure*}

The interaction between magnetic particles depends on their positions and orientations, as defined by their magnetic moment dipoles.
Thus, we also computed the angular pair correlation function $g(r,\theta)$ (Figure~\ref{fig:gr_rt}).
In direct correspondence to the radial pair correlation function $g(r)$ (Figure \ref{fig:gr_r}), brighter regions indicate a higher probability of encountering a pair of particles separated by a distance $r$ at an angle $\theta$ relative to the $z$-axis.
The darker regions at the centers appear because $r$ is the distance between the centers of two particles, and the nanoparticles are hard spheres whose volumes do not overlap. The yellow ring between $r=5$ and $r=10$ for the monodisperse systems is associated with the first neighbor distance, bounded from below by the particle diameter. For polydisperse systems, the minimum distance between two particles' centers can vary significantly therefore, we no longer have such a characteristic first-neighbor-distance ring.

There is a prominent contrast among $g(r,\theta)$ plots of different models. For both poly and monodisperse solutions, the heatmaps from Models 2 and 3 are significantly smoother than those from Model 1. The bright spots in Model 1 (Figures \ref{fig:gr_rt_model_1_monod} and \ref{fig:gr_rt_model_1_polid}) are the result of diverging attractive short-range interactions that trap pairs of nanoparticles in their interaction potential wells. This can also be seen in the tail of the $g(r)$ curves in Figure \ref{fig:gr_r}, although more discreetly due to the averaging over dipole orientations.

\begin{figure*}[ht]
    \begin{center}
    \begin{subfigure}{0.32\hsize}
	\centering
	\includegraphics[width=\linewidth]{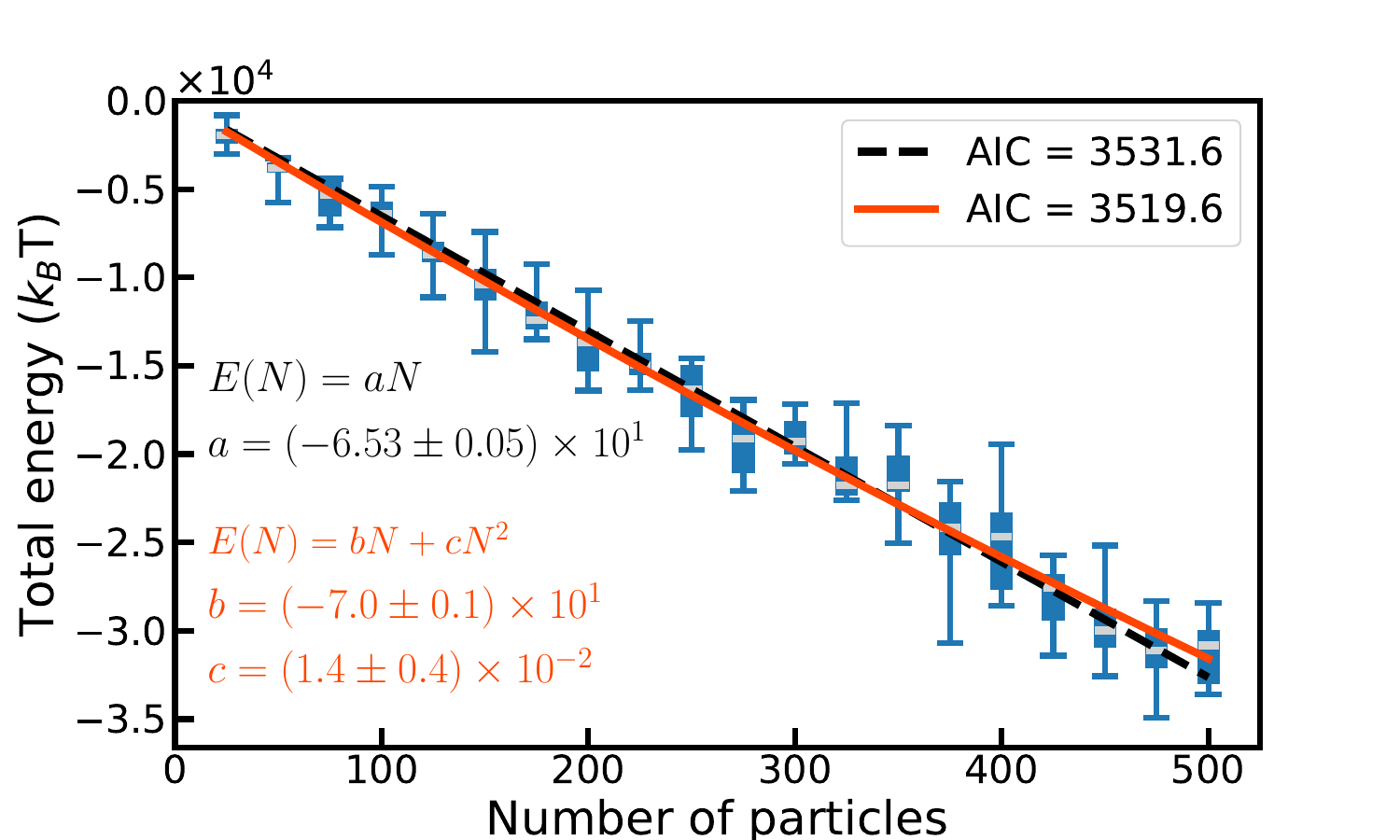}  
	\caption{Model 1, monodisperse}
	\label{fig:u_vs_np_monod_model1}
    \end{subfigure}
    \begin{subfigure}{0.32\hsize}
	\centering
	\includegraphics[width=\linewidth]{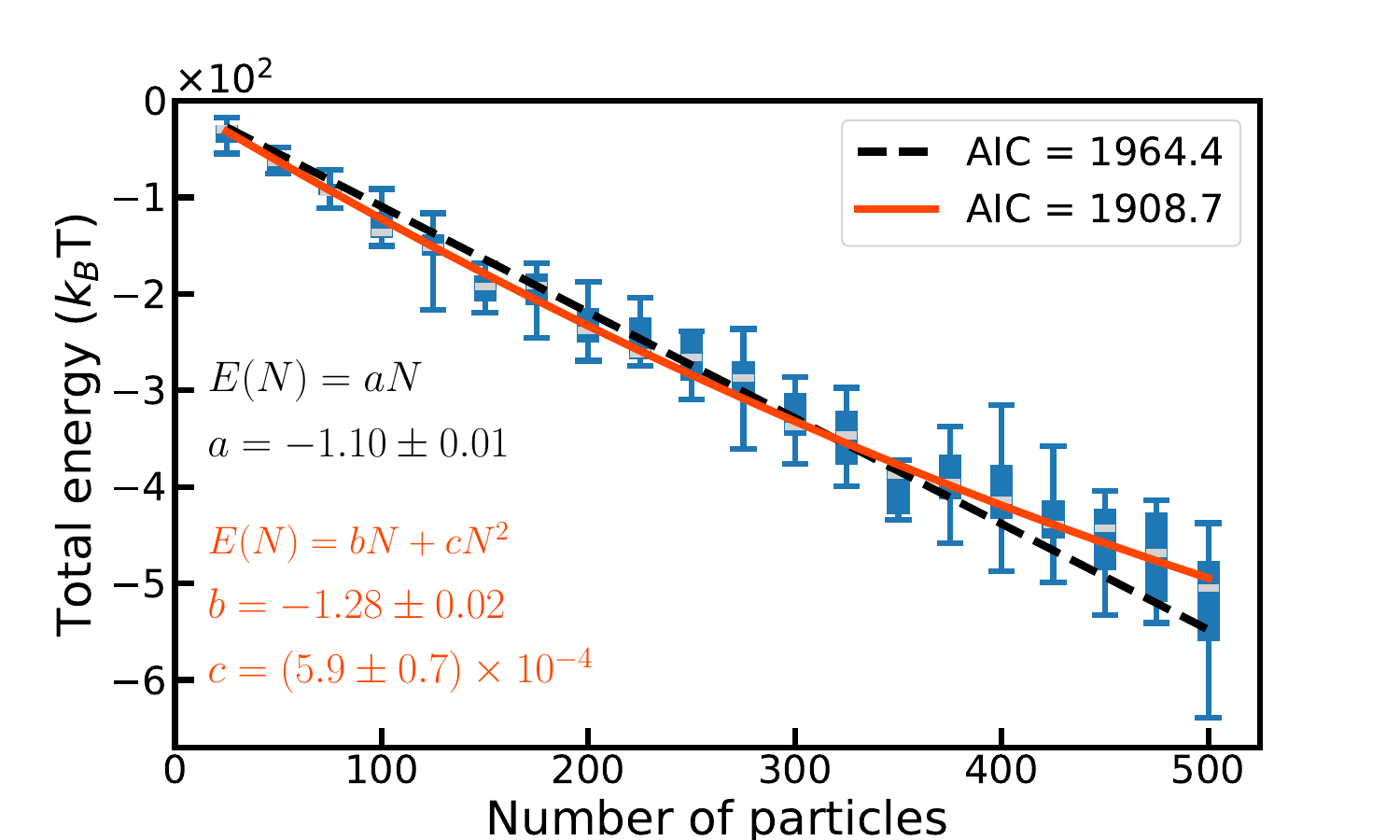}
	\caption{Model 2, monodisperse}
	\label{fig:u_vs_np_monod_model2}
    \end{subfigure}
    \begin{subfigure}{0.32\hsize}
	\centering
	\includegraphics[width=\linewidth]{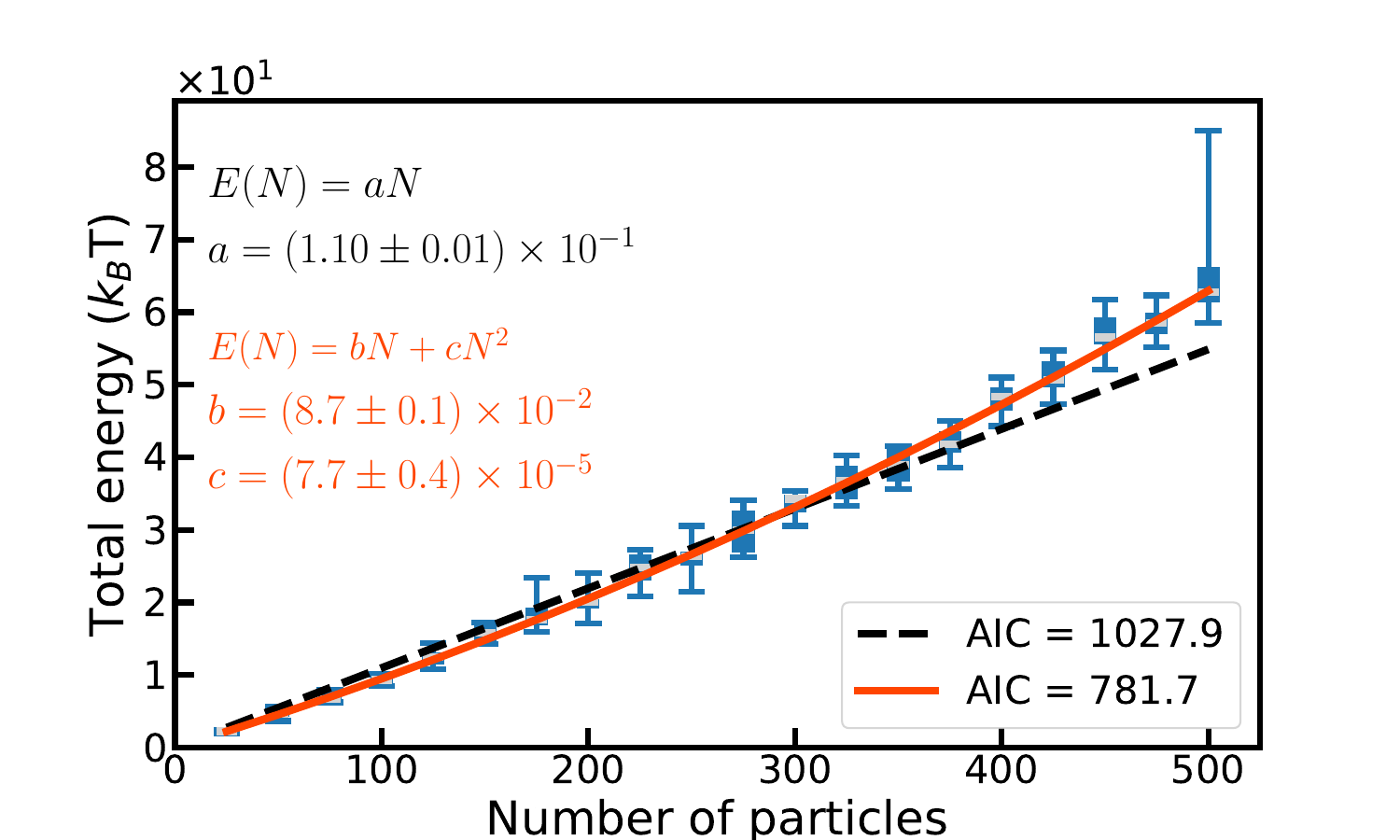}
	\caption{Model 3, monodisperse}
	\label{fig:u_vs_np_monod_model3}
    \end{subfigure}
    \begin{subfigure}{0.32\hsize}
	\centering
	\includegraphics[width=\linewidth]{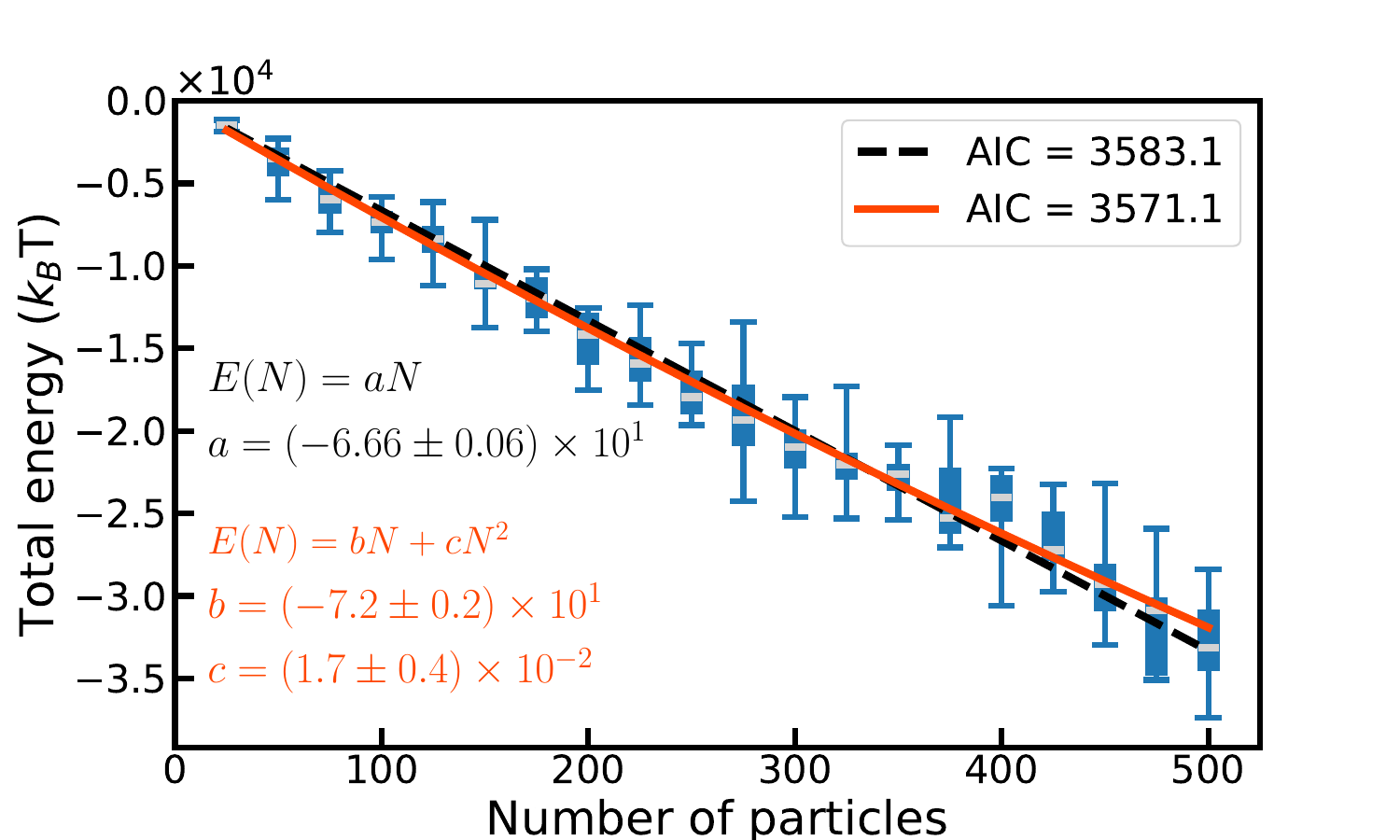}  
	\caption{Model 1, polydisperse}
	\label{fig:u_vs_np_polid_model1}
    \end{subfigure}
    \begin{subfigure}{0.32\hsize}
	\centering
	\includegraphics[width=\linewidth]{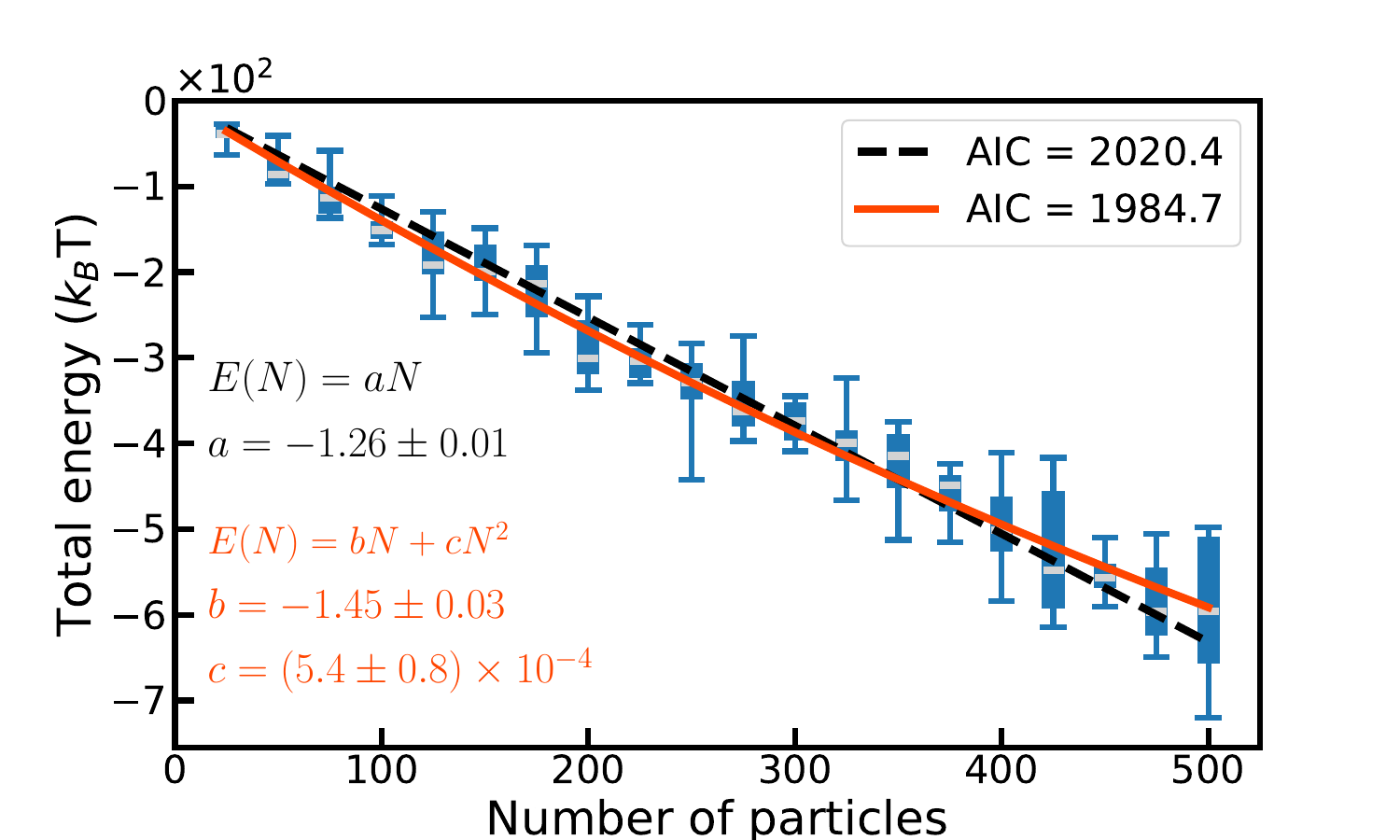}
	\caption{Model 2, polydisperse}
	\label{fig:u_vs_np_polid_model2}
    \end{subfigure}
    \begin{subfigure}{0.32\hsize}
	\centering
	\includegraphics[width=\linewidth]{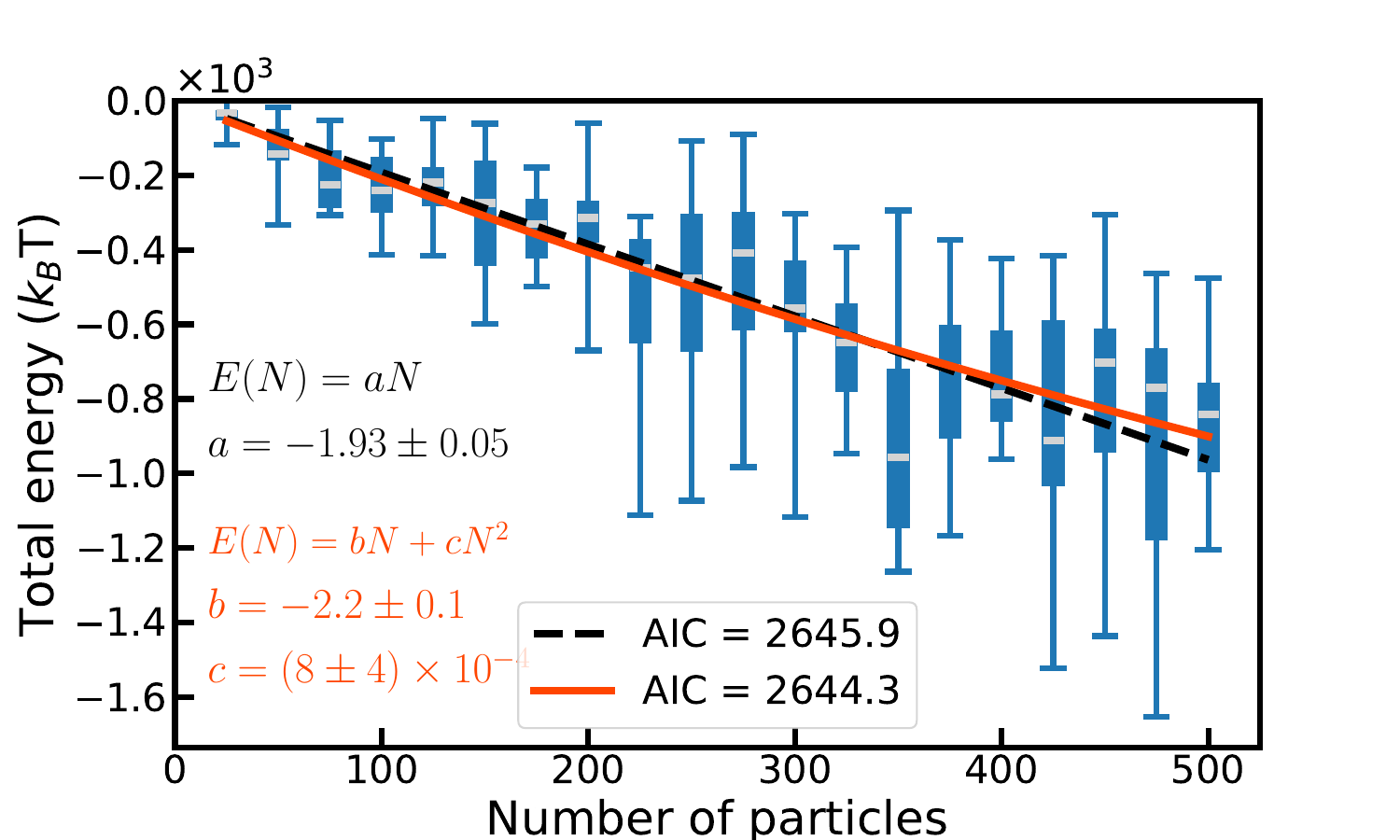}
	\caption{Model 3, polydisperse}
	\label{fig:u_vs_np_polid_model3}
    \end{subfigure}      
    \end{center}
    \caption{\textbf{Magnetic colloids weakly deviate from energy extensivity across all models.} For each model, simulations are done at a volume fraction of $5\%$. In the polydisperse scenario, the median radius is $R_0 = \SI{3.585}{\nano\meter}$, with a log-normal distribution with dispersion $\sigma_R = \num{0.12}$. In the boxplots, the boxes represent the interquartile ranges and the whiskers include the total range of samples, with the grey lines indicating the median values. The black dashed lines show a linear fit of the total energy to particle number, and the solid orange lines show a quadratic fit, in both cases with the intercept set to zero. In all cases, the AIC is smaller for the quadratic fit, confirming the presence of a non-extensive contribution to the energy, albeit relatively small, given estimated values for the parameter $c$ in the quadratic fits.}
    \label{fig:u_vs_np}
\end{figure*}

The highly negative energy values that arise for pairs of particles in Model 1 cause a generally reduced mobility of the particles in the Monte Carlo simulation. When there are interactions with different orders of magnitude, the algorithm prioritizes changes in nanoparticles that most affect the dominant interaction, the diverging van der Waals attraction. As a result, the total energy reaches a deep local minimum (see also the difference of orders of magnitude between the energies of different models in Figure \ref{fig:u_vs_np}). The remaining components lose part of their influence in determining the colloidal stationary configuration. Consequently, finding the global minimum becomes numerically challenging, as changes in the non-dominant interactions produce negligible effects on the total energy. In other words, when the interaction energy between a particle pair is too negative, other particles may also ``freeze'' in energetically unfavorable configurations.

The heatmaps for Models 2 and 3 do not have such prominent spots, presenting only barely discernible lighter regions near the central ring. For Model 2, this result is attributed to the higher EDL repulsion involving larger colloidal particles, which prevents configuration freezing, thus allowing particles to move towards the main minimum of pair interactions.

For Model 3, adopting the Born-Mayer repulsion and the cohesive energy for small separations avoids the implicit assumptions of the van der Waals energy expressions, which commonly do not apply near the microscopic scale. Besides this physical motivation that can provide some trust in these corrections used in Model 3, this model also presents better Monte Carlo sampling properties. Overall, this leads to a higher level of confidence in the results obtained from Model 3.

Our results show that an inadequate application of energy models lead to an unexpected behavior of the simulation algorithm. This can lead to a dramatic bias when estimating physical properties of the system, in particular to the study of agglomeration effects in colloids. Relying on a simplified pair energy function may result in a poor description of the agglomeration mechanism. In contrast, using a more refined approach could not only offer a better representation but also more accurately identify the conditions that lead to these states.

\textbf{Magnetic colloids weakly deviate from energy extensivity across all models.} For the three short-range interaction models, we investigate the energy scaling with the total number of particles in simulations (Figure \ref{fig:u_vs_np}). The energy variation is fit assuming that, for each number $N$ of particles, the energy likelihood follows a normal distribution with fixed $N$-dependent variance $\sigma_{N}^{2}$ and mean given by either a linear model $U_{1}\!\left(N\right) = a N$ or a quadratic one $U_{2}\!\left(N\right) = b N + c N^{2}$. Fitting parameters were obtained via maximum likelihood estimation, solved through a weighted linear regression. Weights were given by the inverses of energies' variances over simulations with the same $N$. We use the implementation of the weighted least squares solver by the Python library \texttt{statsmodels} \citep{Seabold-2010}.

Computing the Akaike information criterion (AIC) \citep{Akaike-1974,Anderson-2004} for the linear and quadratic models, we obtain smaller values for the quadratic model $U_{2}\!\left(N\right)$ in all cases. Therefore, the quadratic model provides a better description of the energy scaling with the number of particles. We do not claim that the true energy scaling is quadratic with the number of particles, though it could be expected given the pairwise nature of interactions. We merely state that our result shows that the energy scaling is nonlinear, as expected given the long range of interactions. 

Notice that the difference in AIC scores is the smallest for Model 3 in the polydisperse case (Fig. \ref{fig:u_vs_np_polid_model3}), given the stronger fluctuation in energy when compared to other models. The enhanced fluctuations reduce the statistical power to identify the contribution of the second-order term, but also indicate higher freedom to reach more diverse configurations with this energy model. Model 3 also contrasts with the other models for giving a positive total energy value for the monodisperse system (Fig. \ref{fig:u_vs_np_monod_model3}), indicating an overall repulsion between the nanoparticles, a clear consequence of the Born-Mayer interaction.
However, for the polydisperse scenario, the total energy is negative as in the other cases, a consequence of the higher magnetic dipolar and van der Waals attractions between the large particles that are present in the polydisperse colloids.

Interestingly, the ratio between the mean values of coefficients $c$ and $b$ is the highest for Model 3 (about an order of magnitude larger than for Models 1 and 2), leading to the largest difference in AIC scores for the monodisperse case. In this regime, the system is near the threshold between the globally repulsive (with overall positive interaction energy) and globally attractive (negative energy) regimes, enhancing the importance of the second-order term $c N^{2}$ -- notice the smaller energy scale in Figure \ref{fig:u_vs_np_monod_model3}.

\section{Discussion}
\label{sec:conclusion}

In summary, we studied three models (Model 1, 2, 3) aimed to simulate magnetic colloids in aggregation regimes. Model 1 represents the standard approach, rooted in the DLVO theory, which served as reference. Model 2 and 3 correspond to proposed corrections in the van der Waals and electric repulsion interactions at the small-range regime.

Our analysis revealed that better short-range interaction models can significantly change predictions about the effect of particle-size dispersity on the colloidal structure. Importantly, the more realistic Model 3 predicts that, at high volume fraction ($\phi = 5 \%$), size polydispersity should decrease the average first neighbor distance $\left\langle s_{1} \right\rangle$, opposite to what is expected from Models 1 and 2. This difference could provide an interesting direction for future experimental research. Experimentally identifying more realistic models for describing magnetic colloids could significantly impact the development of the field. It would allow for more trustworthy theoretical predictions about costly and hard-to-perform experiments, such as studying magnetic colloids in complex environments.

Moreover, improving short-range interaction models also aids convergence of Monte Carlo simulations by reducing sampling biases. The main problem with the standard DLVO theory, characterized here by Model 1, is the divergence of the van der Waals energy to negative infinity as the distance between the surfaces of two particles tends to zero. This divergence dramatically hampers the Monte Carlo sampling of the system, as seen in the salt and pepper patterns in the heatmaps for Model 1 in Figure \ref{fig:gr_rt}. This happens due to van der Waals energy contribution that leads to deep energy wells at very short distances. Consequently, many particle pairs fall into these local minima, unrealistically reducing the overall mobility of particles during the simulations. 

We provided two different models that solve the energy divergence, improving Monte Carlo sampling compared to the conventional DLVO approach \citep{tufenkji2004deviation,Israelachvili-1992,hoek2003effect,PAULA-2007, Paula-2009,russel1991colloidal,shen2012application}. In particular, Model 3 is more realistic from a theoretical point of view, as it substitutes the van der Waals attraction with the cohesive energy and the Born-Mayer repulsion at short distances. We argue that Model 3 might have better sampling properties than Model 2: For all 3 models, the secondary energy minimum moves to smaller distances as we shift from monodisperse to polydisperse samples (Supplementary Figures S3 and S4). However, we only see a decrease in the average first neighbor distance $\left\langle s_{1} \right\rangle$ due to polydispersity in simulations of Model 3. This could be an indication of better sampling properties of Model 3 compared to Model 2. We still need experimental data to confirm the correctness of this prediction, as mentioned above.

As long-range interactions can lead to non-extensive energies, we investigated the change in total energy with the number of particles for a high-density ferrofluid solution, and its variation across models.
The results showed a slight shift from linearity for all models studied, confirmed by the AIC scores of linear and nonlinear models of energy scaling with the number of particles. The nonlinearity was more pronounced for Model 3 in monodisperse configuration. This is a borderline case between the globally repulsive and attractive regimes, leading to smaller absolute energy values. Provided experimental validation, the models can be useful for precise identification of ideal ideal particle size to maintain the necessary colloidal structure of different applications.

\section*{Supporting Information}

Supplementary files associated with this article can be found in the online version.

\section*{Acknowledgements}

The authors gratefully acknowledge financial support from Fundação de Apoio à Pesquisa do Distrito Federal (FAPDF) through Grant No. 00193-00001713/2024-10.

\FloatBarrier

\appendix

\section{Parameters}
\label{sec:appendix:parameters}

    \begin{table}[H]
        \centering
        \caption{Metropolis-Hasting algorithm parameters.}
	\begin{tabular}{|p{0.1\hsize}|p{0.35\hsize}|p{0.35\hsize}|}
           \hline
		\multicolumn{1}{|c|}{\textbf{Variable}}	&	\multicolumn{1}{|c|}{\textbf{Description}}	&	\multicolumn{1}{|c|}{\textbf{Value}}	\\
		\hline
		$N_{MC[max]}$		&	maximum number of MC steps	&	\makecell[c]{$\num{300000}$}	\\
		\hline
		$N_{MC[chk]}$		&	number of MC steps between convergence checks	&	\makecell[c]{$\num{50000}$}	\\
		\hline
		$\Delta{U}_{ini}$	&	energy variation to begin sampling	&	\makecell[c]{$\num{0.05}$}	\\
		\hline
		$\Delta{U}_{fin}$	&	energy variation to finish sampling &	\makecell[c]{$\num{0.005}$}	\\
		\hline
		$\alpha$			&	target acceptance rate in Metropolis algorithm				&	\makecell[c]{$\num{0.5}$}	\\
		\hline
  		$s_{min}$		&	surface minimum distance &	\makecell[c]{$\SI{0.01}{\nano\meter}$}				\\
		\hline
  		$N$	&	number of particles in the box  &	\makecell[c]{$\num{25}$	(minimum) \\ $\num{500}$ (maximum)} \hfill \citep{Bakuzis-2013}  	\\
		\hline
	\end{tabular}	
    \end{table}

    \begin{table}[H]
	\centering
        \caption{Dispersed phase: magnetite nanoparticles.}
		\begin{tabular}{|p{0.1\hsize}|p{0.3\hsize}|p{0.35\hsize}|}
			\hline
			\multicolumn{1}{|c}{\textbf{Parameters}}	&	\multicolumn{1}{|c}{\textbf{Description}}	&	\multicolumn{1}{|c|}{\textbf{Value}}		\\
		\hline
		\makecell[c]{$D_0$ \\ $R_0$}		&	\makecell[l]{median diameter \\ median radius}	&	\makecell[c]{$\SI{7.17}{\nano\meter}$ \\ $\SI{3.585}{\nano\meter}$} \hfill \citep{Bakuzis-2013}		\\
		\hline
		\makecell[c]{$\sigma_D$ \\ $\sigma_R$}	&	\makecell[l]{diameter dispersion \\ radius dispersion}	&	\makecell[c]{$\num{0.24}$ \\ $\num{0.24}$} \hfill \citep{Bakuzis-2013}\\
		\hline
		\makecell[c]{$M$}			&	magnetization	&	\makecell[c]{$\SI{4.71e5}{\ampere\per\meter}$} \hfill \citep{Bakuzis-2013}\\
		\hline
		\makecell[c]{$\xi$}		&	grafting parameter 	&	\makecell[c]{$\SI{2.0e17}{\per\meter\squared}$} \hfill \citep{Bakuzis-2013}	\\
		\hline
		\makecell[c]{$\delta$}	&	surfactant layer length  &	\makecell[c]{$\SI{0.55}{\nano\meter}$} \hfill \citep{Bakuzis-2013}\\
		\hline
		\makecell[c]{$\delta_s$}	&	excluded radius		&	\makecell[c]{$\SI{0.8}{\nano\meter}$} \hfill		\\
		\hline
		\makecell[c]{$W_{coh}$} 	&	cohesive energy (maghemite)	&	\makecell[c]{$\SI{1.38}{\joule\per\meter\squared}$} \hfill	\citep{Diakonov-1998}\\
		\hline
		\makecell[c]{$L_{B}$} 	&	atomic bond distance  (maghemite \ce{Fe-O}) &	\makecell[c]{$\SI{0.19}{\nano\meter}$} \hfill \citep{Fdez-2016, Coduri-2020}	\\
		\hline	
		\makecell[c]{$A_{bm}$}		&	Born-Mayer parameter A	&	\makecell[c]{$\SI{1207.6}{\electronvolt}$} \hfill \citep{Lewis-1985}		\\
		\hline
		\makecell[c]{$L_{bm}$}		&	Born-Mayer parameter L	&	\makecell[c]{$\SI{0.3084}{\angstrom}$} \hfill \citep{Lewis-1985}		\\
		\hline
		\makecell[c]{$\lambda_D$}		&	Debye length			&	\makecell[c]{$\SI{1.1}{\nano\meter}$} \hfill \citep{Bakuzis-2013}\\
		\hline
	\end{tabular}	
    \end{table}

    \begin{table}[H]
	\centering
        \caption{Dispersion medium: water.}
	\begin{tabular}{|p{0.1\hsize}|p{0.3\hsize}|p{0.35\hsize}|}
			\hline
			\multicolumn{1}{|c}{\textbf{Parameters}}	&	\multicolumn{1}{|c}{\textbf{Description}}	&	\multicolumn{1}{|c|}{\textbf{Value}}		\\
			\hline
			\makecell[c]{$\mu$}		&	 magnetic permeability	&	\makecell[c]{$\SI{1.26e-6}{\henry\per\meter}$} \hfill	\\
			\hline
			\makecell[c]{$\rho_{ion}$}	&	ion concentration		&	\makecell[c]{$\SI{0.15}{\mol\per\litre}$} \hfill \citep{Bakuzis-2013}	\\
			\hline
	\end{tabular}
    \end{table}

    \begin{table}[H]
	\centering
        \caption{System's parameters.}
	\begin{tabular}{|p{0.1\hsize}|p{0.3\hsize}|p{0.35\hsize}|}
			\hline
			\multicolumn{1}{|c}{\textbf{Parameters}}	&	\multicolumn{1}{|c}{\textbf{Description}}	&	\multicolumn{1}{|c|}{\textbf{Value}}		\\
                \hline
   			\makecell[c]{$\phi$}			&	volume fraction 		&	\makecell[c]{$\num{0.05}$ ($\SI{5}{\percent}$) \\ $\num{0.0047}$  ($\SI{0.47}{\percent}$)} \hfill \citep{Bakuzis-2013}\\
			\hline
			\makecell[c]{$T$}	&	temperature			&	\makecell[c]{$\SI{300}{\kelvin}$} \hfill \citep{Bakuzis-2013}	\\
			\hline
			\makecell[c]{$A$}			&	Hamaker constant 		&	\makecell[c]{$\SI{0.40e-21}{J}$} \hfill \citep{Bakuzis-2013}\\
			\hline
	\end{tabular}	
    \end{table}

 

 \bibliographystyle{ieeetr}






\end{document}